\documentclass[%
 aip,
 jmp,%
 amsmath,amssymb,
 reprint,%
author-year,%
]{revtex4-2}

\usepackage{graphicx}% Include figure files
\usepackage{dcolumn}% Align table columns on decimal point
\usepackage{bm}% bold math
\usepackage[]{float}
\usepackage{amsmath}
\usepackage{tabularx}
\usepackage{booktabs}
\usepackage{multirow}
\usepackage{hyperref}
\usepackage{amssymb}
\usepackage{mathtools}
\usepackage{changepage}
\usepackage{accents}

\usepackage[table]{xcolor}

\newcommand{\ubar}[1]{\underaccent{\bar}{#1}}

\newlength{\fulllength}
\newlength{\extralength}
\begin{document}

%\preprint{AIP/123-QED}

\title{Runtime Construction of Large-Scale Spiking Neuronal Network Models on GPU Devices}

\author{Bruno Golosio}
\altaffiliation{Equal contribution}
\affiliation{Department of Physics, University of Cagliari, Italy}
\affiliation{Istituto Nazionale di Fisica Nucleare (INFN), Sezione di Cagliari, Cagliari, Italy}

\author{Jose Villamar}
\altaffiliation{Equal contribution}
\affiliation{Institute of Neuroscience and Medicine (INM-6), Institute for Advanced Simulation (IAS-6), JARA-Institute Brain Structure-Function Relationships (INM-10), Jülich Research Centre, Jülich, Germany}

\author{Gianmarco Tiddia}
\altaffiliation{Equal contribution}
\email[Correspondence:]{gianmarco.tiddia@dsf.unica.it}
\affiliation{Department of Physics, University of Cagliari, Italy}
\affiliation{Istituto Nazionale di Fisica Nucleare (INFN), Sezione di Cagliari, Cagliari, Italy}

\author{Elena Pastorelli}
\affiliation{Istituto Nazionale di Fisica Nucleare, Sezione di Roma, Italy}

\author{Jonas Stapmanns}
\affiliation{Institute of Neuroscience and Medicine (INM-6), Institute for Advanced Simulation (IAS-6), JARA-Institute Brain Structure-Function Relationships (INM-10), Jülich Research Centre, Jülich, Germany}

\author{Viviana Fanti}
\affiliation{Department of Physics, University of Cagliari, Italy}
\affiliation{Istituto Nazionale di Fisica Nucleare (INFN), Sezione di Cagliari, Cagliari, Italy}

\author{Pier Stanislao Paolucci}
\affiliation{Istituto Nazionale di Fisica Nucleare, Sezione di Roma, Italy}

\author{Abigail Morrison}
\affiliation{Institute of Neuroscience and Medicine (INM-6), Institute for Advanced Simulation (IAS-6), JARA-Institute Brain Structure-Function Relationships (INM-10), Jülich Research Centre, Jülich, Germany}
\affiliation{Department of Computer Science 3 - Software Engineering, RWTH Aachen University, Aachen, Germany}

\author{Johanna Senk}
\affiliation{Institute of Neuroscience and Medicine (INM-6), Institute for Advanced Simulation (IAS-6), JARA-Institute Brain Structure-Function Relationships (INM-10), Jülich Research Centre, Jülich, Germany}

\date{\today}% It is always \today, today,
             %  but any date may be explicitly specified

\begin{abstract}
Simulation speed matters for neuroscientific research: this includes not only how quickly the simulated model time of a large-scale spiking neuronal network progresses, but also how long it takes to instantiate the network model in computer memory.
On the hardware side, acceleration via highly parallel GPUs is being increasingly utilized.
On the software side, code generation approaches ensure highly optimized code, at the expense of repeated code regeneration and recompilation after modifications to the network model.
%(2) Methods
Aiming for a greater flexibility with respect to iterative model changes, here we propose a new method for creating network connections interactively, dynamically, and directly in GPU memory through a set of commonly used high-level connection rules.
%(3) Results
We validate the simulation performance with both consumer and data center GPUs on two neuroscientifically relevant models:
a cortical microcircuit of about $77,000$ leaky-integrate-and-fire neuron models and $300$ million static synapses, and a two-population network recurrently connected using a variety of connection rules.
%(4) Conclusions
With our proposed ad hoc network instantiation, both network construction and simulation times are comparable or shorter than those obtained with other state-of-the-art simulation technologies, while still meeting the flexibility demands of explorative network modeling.
\end{abstract}

%Use showkeys class option if keyword
                              %display desired
\maketitle
\section{Introduction}
\label{sec:intro}
Spiking neuronal network models are widely used in the context of computational neuroscience to study the activity of populations of neurons in the biological brain. Numerous software packages have been developed to simulate these models effectively. Some of these simulation engines offer the ability to accurately simulate a wide range of neuron models and their synaptic connections. Among the most popular codes are NEST \citep{Gewaltig_07_11204}, NEURON \citep{Carnevale_2006}, Brian 2 \citep{Stimberg2019}. NENGO \citep{Bekolay2014} and ANNarchy \citep{Vitay2015} should also be mentioned. In recent years there has been a growing interest in GPU-based approaches, which can be particularly useful for simulating large-scale networks thanks to their high degree of parallelism. This interest is also fueled by the rapid technological development of this type of device and by the availability of increasingly performant GPU cards, both for the consumer and for high-performance computing (HPC) infrastructure.
A main driving force behind this development is the demand from current artificial intelligence algorithms and similar applications for massively parallel processing of simple floating point operations, and a corresponding industry with huge financial resources.
Present day supercomputers are reaching for exascale by drawing their compute power from GPUs.
For neuroscience to benefit from these systems, efficient algorithms for the simulation of spiking neuronal networks need to be developed.
Simulation codes such as GeNN \citep{Yavuz2016}, CARLsim \citep{Nageswaran2009,CARLsim6}, and NEST GPU \citep{Golosio2021} have been primarily designed for GPUs, while in recent times, popular CPU-based simulators have shown interest in integrating the more traditional CPU-based approach with libraries for GPU simulation \citep{Kumbhar2019, Golosio2020, Stimberg2020, Tiddia2022, Alevi2022, Awile2022}. Also the novel simulation library Arbor \citep{paper:arbor2019}, which focuses on morphologically-detailed neural networks, takes GPUs into account.\\
In general, GPU-based simulators fall into one of three categories: those that allow the construction of network models at run time using scripting languages, those that require the network models to be fully specified in a compiled language, and hybrid ones that provide both options. The most extensively used compiled languages are C and C++ for host code, and CUDA for device code (using NVIDIA GPUs) while the most widely used scripting language is Python. With scripting languages, simulations can be performed without the need to compile the code used to describe the model. Consequently, the time required for compilation is eliminated. Furthermore, in many cases the use of a scripting language simplifies the implementation of the model, especially for users who do not have extensive programming language expertise.
Approaches using compiled languages typically have much faster network construction times. To reconcile this benefit with the greater ease of model implementation using a scripting language, some simulators have shifted toward a code-generation approach \citep{Vitay2015, Yavuz2016}. In this approach, the model is implemented by the user through a brief high-level description, which the code generator then converts into the language or languages that must be compiled before being executed by the CPU and GPU. The main disadvantage of code-generation based simulators is the need for new code generation and compilation every time model modifications such as changes in network architecture are necessary. The times associated with code generation and compilation are typically much longer than network construction times \citep{Golosio2020}.\\
Examples of the code-generation based approaches include GeNN \citep{Yavuz2016} and ANNarchy \citep{Vitay2015}. In GeNN, neuron and synapse models are defined in C++ classes, and snippets of C-like code can be used to offload costly operations onto the GPU. The Python package PyGeNN \citep{Knight2021_pygenn} is built on top of an automatically generated wrapper for the C++ interface (using SWIG\footnote{\url{https://www.swig.org}}) and allows for the same low-level control.
Further, Brian2GeNN \citep{Stimberg2020} provides a code generation pipeline for defining models via the Python interface of Brian \citep{Stimberg2019} and using GeNN as a simulator backend. Alternatively, Brian2CUDA \citep{Alevi2022} directly extends Brian with a GPU backend.
The hybrid approach is exemplified by CARLsim \citep{CARLsim6}, which has also developed its own Python interface to communicate with its C/C++ library, named PyCARL \citep{PyCarl}. Much like PyNEST \citep{PyNEST}, the Python interface of the NEST simulator, CARLsim exposes its C/C++ kernel through a dynamic library which can then interact with Python, however like GeNN they make use of SWIG to automatically generate the binding between their library and their Python interface. PyCARL directly serves as a PyNN \citep{Davison2008} interface.
NEST GPU \citep{Tiddia2022} is a software library for the simulation of spiking neural networks on GPUs. It originates from the prototype NeuronGPU library \citep{Golosio2021} and is now overseen by the NEST Initiative and integrated with the NEST development process. NEST GPU uses a hybrid approach and offers the possibility to implement models using either Python scripts or C++ code. The main commands of the Python interface, the use of dictionaries, the names and parameters of the neuron and spike generator models, are already aligned to those of the CPU-based NEST code. In previous version of NEST GPU, connections were first created on the CPU side and then copied from RAM to GPU memory. This approach benefited from the standard C++ libraries, in particular the dynamic allocation of container classes of the C++ Standard Template Library, and used a multi-threaded approach on the CPU via the OpenMP library.
However, it had the drawback of relatively long network construction times, not only due to the costly copying of connections and other CPU-side initializations, but also when CPUs with a limited number of cores were used, restricting the level of parallelization for creating the connections.\\
This work proposes a network construction method in which the connections are created directly in the GPU memory with a dynamic approach, and then suitably organized in the same memory using algorithms that exploit GPU parallelism. This approach, so far applied to single-GPU simulations, enables much faster connection creation, initialization and organization, while preserving the advantages of dynamic connection building, particularly the ability to create and initialize the model at run-time without the need for compilation. Although this method was developed specifically in the framework of NEST GPU, the concepts are sufficiently general that they should be applicable with minimal adaptation to other GPU-based simulators, as far as they are designed with a modular structure.\\
The Materials and Methods section of this manuscript first introduces the dynamic creation of connections and provides details on the used data structures and the spike buffer employed by the simulation algorithm (Sections \ref{sec:methods_dyn_conns} -- \ref{sec:methods_spike_buffer}); details on the employed block sorting algorithm are in Appendix \ref{sec:appendix_block_sorting}. The proposed dynamic approach for network construction is tested on the simulation of two complementary network models across different hardware configurations; we then compare the performance to other simulation approaches.
Details on the network models, the hardware and software, and time measurements for performance evaluation are given in Sections \ref{sec:methods_models} -- \ref{sec:methods_phases}. The spiking activity of a network constructed with the dynamic approach is validated statistically in Section \ref{sec:methods_validation} and Appendix \ref{sec:appendix_validation}.
The performance results are shown in Section \ref{sec:results} (with additional data in Appendices \ref{sec:appendix_microcircuit} and \ref{sec:appendix_izhikevich}) and discussed in Section \ref{sec:discussion}.

\section{Materials and Methods}
\label{sec:methods}
\subsection{Creation of connections directly in GPU memory}
\label{sec:methods_dyn_conns}

A network model is composed of nodes, which are uniquely identifiable by index and connections between them. In NEST and NEST GPU, a node can be either a neuron or a device for stimulation or recording.
Neuron models can have multiple receptor ports to allow receptor-specific parameterization of input connections.
Connections are defined in NEST GPU (and similarly in other simulators) via high-level connection routines, e.g., \texttt{ngpu.Connect(sources, targets, conn\_dict, syn\_dict)}, where the connection dictionary \texttt{conn\_dict} specifies a connection rule, e.g., \texttt{one\_to\_one}, for establishing connections between source and target nodes.
The successive creation of several individual sub-networks, according to deterministic or probabilistic rules, can then lead to a complex overall network.
In the rules used here, we allow autapses (self-connections) and multapses (multiple connections between the same pair of nodes); see \cite{Senk2022} for a summary of state-of-the-art connectivity concepts.

The basic structure of a NEST GPU connection includes the source node index, the target node index, the receptor port index, the weight, the delay, and the synaptic group index.
The synaptic group index takes non-zero values only for non-static synapses (e.g., STDP) and refers to a structure used to store the synapse type and the parameters common to all connections of that group; it should not be confused with the connection group index, which groups connections with the same delay for spike delivery.
Additional parameters may be present depending on the type of synapse, which is specified by the synaptic group. The delay must be a positive multiple of the simulation time resolution, and can therefore be represented using time-step units as a positive integer.
Connections are stored in GPU memory in dynamically-allocated blocks with a fixed number of connections per block, $B$, which can be specified by the user as a simulation kernel parameter before creating the connections. It should be chosen on the basis of a compromise. If it is chosen too small, then the total number of blocks would be high, resulting in larger execution times. Conversely, if it is chosen too large, a significant amount of memory could be wasted due to incomplete filling of the last allocated block.
The default value for $B$ used in all simulations of this study is $10^7$.

\begin{figure}[t]
%\begin{adjustwidth}{-\extralength}{0cm}
\centering
\includegraphics[width=\textwidth]{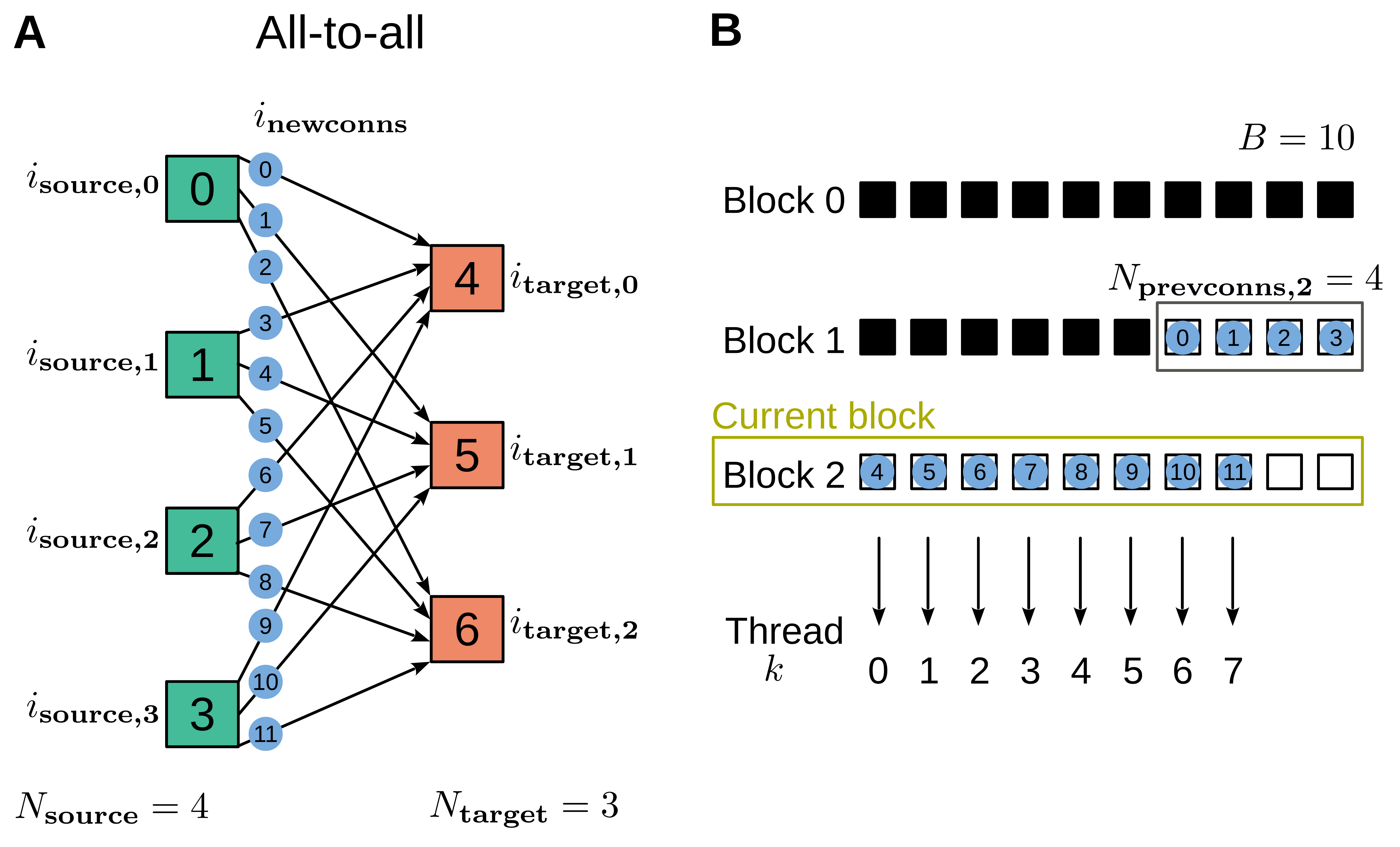}
%\end{adjustwidth}
\caption{
Example of connection creation through the all-to-all connection rule.
\textbf{(A)}~Each one of the four source nodes (green) is connected to all three target nodes (orange).
The connections generated by this rule are identified by an index, $i_\textbf{newconns}$, which here ranges from $0$ to $11$ (blue disks).
\textbf{(B)}~The connections are stored in blocks that are allocated dynamically, where for demonstration purposes a block size of ten connections is used.
The black squares represent previous connections (established through an earlier connect call), while the twelve connections generated by the considered instance of the all-to-all rule are represented by the same blue disks labeled with $i_\textbf{newconns}$ as in panel A.
The new connections in different blocks are generated by separate CUDA kernels.
In this example, $N_\textbf{prevconns,2}$ of the new connections are created in the previous block (grey frame), and the remaining ones in the current block ($b=2$, yellow frame), where $i_\textbf{newconns}$ is computed by adding the CUDA thread index $k$ to $N_\textbf{prevconns,2}$.
\label{fig:conn_sketch}}
\end{figure}

Each time a new connection-creation command is launched, if the last allocated block does not have sufficient free slots to store the new connections, an appropriate number of new blocks is allocated, according to the formula:
\begin{equation}
N_\textbf{newblocks} = \lfloor  \frac {N_\textbf{conns} + N_\textbf{newconns} + B - 1} {B} \rfloor - N_\textbf{blocks}
\end{equation}
where $N_\textbf{newblocks}$ is the number of new blocks that must be allocated, $N_\textbf{blocks}$ is the old number of blocks, $N_\textbf{conns}$ is the old number of connections, $N_\textbf{newconns}$ is the number of connections that must be created, and $\lfloor x \rfloor$ denotes the integer part of $x$.
The new connections are indexed contiguously:
\begin{equation}
    i_\textbf{newconns} = 0, ..., N_\textbf{newconns}-1.
\end{equation}
A loop is performed on the blocks, starting from the first block in which there are available slots up to the last allocated block, and the connections are created in each block by launching appropriate CUDA kernels\footnote{CUDA kernels are functions executed on the GPU device. These kernels concurrently exploit multiple CUDA-thread blocks.} to set the connection parameters described above.
In each block $b$, the index of each of the new connections is calculated from the CUDA-thread\footnote{CUDA-threads are the smallest GPU computing units. These threads are grouped into blocks and several blocks are present in a multiprocessor unit.} index $k$ according to the formula:
\begin{equation}
i_\textbf{newconns,b} = N_\textbf{prevconns,b} + k\quad \text{with}\quad k=0,...,N_\textbf{thr} -1
\end{equation}
where $i_\textbf{newconns,b}$ refers to the subset of $i_\textbf{newconns}$ on the current block and $N_\textbf{prevconns,b}$ is the number of new connections created in the previous blocks.
The number of connections to be created in the current block, which corresponds to the number of required threads $N_\textbf{thr} $, is computed before launching the kernel; if the block will be completely filled, the number of threads equals the block size, $N_\textbf{thr} =B$.
See Figure \ref{fig:conn_sketch} for an example of how the connections are numbered and assigned to the blocks.\\

The indexes of a source node $s$ and a target node $t$ are calculated from $i_\textbf{newconns}$ using expressions that depend on the connection rule. Here we provide both the name of the rules as defined in \cite{Senk2022} and their corresponding parameter of the NEST interface. In case that both the source-node group and the target-node group contain nodes with consecutive indexes, starting from $s_0$ and from $t_0$, respectively, the node indexes are:
\begin{itemize}
\setlength\itemsep{0.5em}
    \item \textbf{One-to-one} (\texttt{one\_to\_one}):
\begin{align}
s &= s_0 + i_\textbf{newconns} \\
t &= t_0 + i_\textbf{newconns}
\end{align}
with $N_\textbf{newconns} = N_\textbf{sources} = N_\textbf{targets}$.

    \item \textbf{All-to-all} (\texttt{all\_to\_all}):
\begin{align}
s &= s_0 + \lfloor \frac{i_\textbf{newconns}}  {N_\textbf{targets}} \rfloor \\
t &= t_0 +  \textbf{mod}(i_\textbf{newconns}, N_\textbf{targets})
\end{align}
with $N_\textbf{newconns} = N_\textbf{sources} \times N_\textbf{targets}$.

    \item \textbf{Random, fixed out-degree with multapses} (\texttt{fixed\_outdegree}):
\begin{align}
s &= s_0 + \lfloor \frac{i_\textbf{newconns}}  {K} \rfloor \\
t &= t_0 +  \textbf{rand}(N_\textbf{targets})
\end{align}
where $K$ is the out-degree, i.e., the number of output connections per source node, $\textbf{rand}(N_\textbf{targets})$ is a random integer between 0 and $N_\textbf{targets}-1$ sampled from a uniform distribution, and $N_\textbf{newconns} = N_\textbf{sources} \times K$.

    \item \textbf{Random, fixed in-degree with multapses} (\texttt{fixed\_indegree}):
\begin{align}
s &= s_0 + \textbf{rand}(N_\textbf{sources}) \\
t &= t_0 +  \lfloor \frac{i_\textbf{newconns}}  {K} \rfloor
\end{align}
where $K$ is the in-degree, i.e., the number of input connections per target node, and $N_\textbf{newconns} = n_\textbf{targets} \times K$.

    \item \textbf{Random, fixed total number with multapses} (\texttt{fixed\_total\_number}):
\begin{align}
s &= s_0 + \textbf{rand}(N_\textbf{sources}) \\
t &= t_0 +  \textbf{rand}(N_\textbf{targets})
\end{align}
where pairs of sources and targets are sampled until the specified total number of connections $N_\textbf{newconns}$ is reached.

\end{itemize}
If the indexes of source or target nodes are not consecutive but are explicitly given by an array, the above formulas are used to derive the indexes of the array elements from which to extract the node indexes.
Weights and delays can have identical values for all connections, or be specified for each connection by an array having a size equal to the number of connections, or be randomly distributed according to a given probability distribution.
In the latter case, the pseudo-random numbers are generated using the cuRAND library\footnote{\url{https://developer.nvidia.com/curand}}.
The delays are then converted to integer numbers expressed in units of the computation time step by dividing their values, expressed in milliseconds, by the duration of the computation time step, and rounding the result to an integer. The minimal delay that is permitted is one computation time step \citep{Morrison08_267}, thus if the result is less than $1$, the delay is set to $1$ in time step units.
\subsection{Data structures used for connections}
\label{sec:methods_orga_conns}
In order to efficiently manage the spike transmission in the presence of delays, the connections must be organized in an appropriate way. To this end, the algorithm divides the connections into groups, so that connections from the same group share the same source node and the same delay. This arrangement is needed for the spike delivery algorithm, which is described in the next section. The algorithm  achieves this by hierarchically using two sorting keys: the index of the source node as the first key and the delay as the second. Since the connections are created dynamically, their initial order is arbitrary. Therefore we order connections in a stage that follows network construction and that precedes the simulation, called \textit{calibration} phase (for a definition of the simulation phases see Section \ref{sec:methods_phases}). The sorting algorithm is an extension of radix-sort \citep{Cormen2009} applied to an array organized in blocks, based on the implementation available in the CUB library\footnote{\url{https://nvlabs.github.io/cub}}.
Once the connections are sorted, their groups must be adequately indexed, so that when a neuron emits a spike, the code has quick access to the groups of connections outgoing from this neuron and to their delays.
This indexing is done in parallel using CUDA kernels on connection blocks with one CUDA thread for each connection. The connection index extracts the source node index and the connection delay. If one of these two values differs from those of the previous connection, it means that the current connection is the first of a connection group. We use this criterion to count the number of connection groups per source node, $G_i$, and to find the position of each connection group in the connection blocks.
The next step constructs for each source node an array of size equal to the number of groups of outgoing connections containing the global indexes of the first connections of each group. Since allocating a separate array for each node would be a time-consuming operation, we concatenate all arrays into a single one-dimensional array. The starting position $p_i$ of the sub-array corresponding to a given source node $i$ can be evaluated by the cumulative sum of $G_i$ as follows
\begin{equation}
  p_i = \sum_{j=0}^{i-1}{G_j} \qquad i=1, \ldots, N_\textbf{nodes} \qquad \text{and} \qquad p_0=0
\end{equation}
where $N_\textbf{nodes}$ is the total number of nodes in the network.

\subsection{The spike buffer}
\label{sec:methods_spike_buffer}
The simulation algorithm  employs a buffer of outgoing spikes for each neuron in the network to manage connection delays \citep{Golosio2021, Tiddia2022}. Each spike object is composed of three parameters: a time index, a connection group index and a multiplicity (i.e., the number of physical spikes emitted by a network node in a single time step). The spike buffer has the structure of a queue into which the spikes emitted by the neuron are inserted. Whenever a spike is emitted from the neuron, it is buffered, and both its time index and its connection-group index are initialized to zero. At each simulation time step, the time indexes of all the spikes are increased by one unit. When the time index of a spike matches the delay of the connection group indicated by its connection group index, the spike is fed into a global array called \textit{spike array}, and its connection group index is incremented by one unit, so as to point to the next connection group in terms of delay. In the spike array, each spike is represented by the source node index, the connection group index and the multiplicity. The spikes are delivered in parallel to the target nodes using a CUDA kernel with one CUDA thread for each connection of each connection group inserted in the spike array.

\subsection{Models used for performance evaluation}
\label{sec:methods_models}

\begin{figure}[t]
\centering
\includegraphics[width=\textwidth]{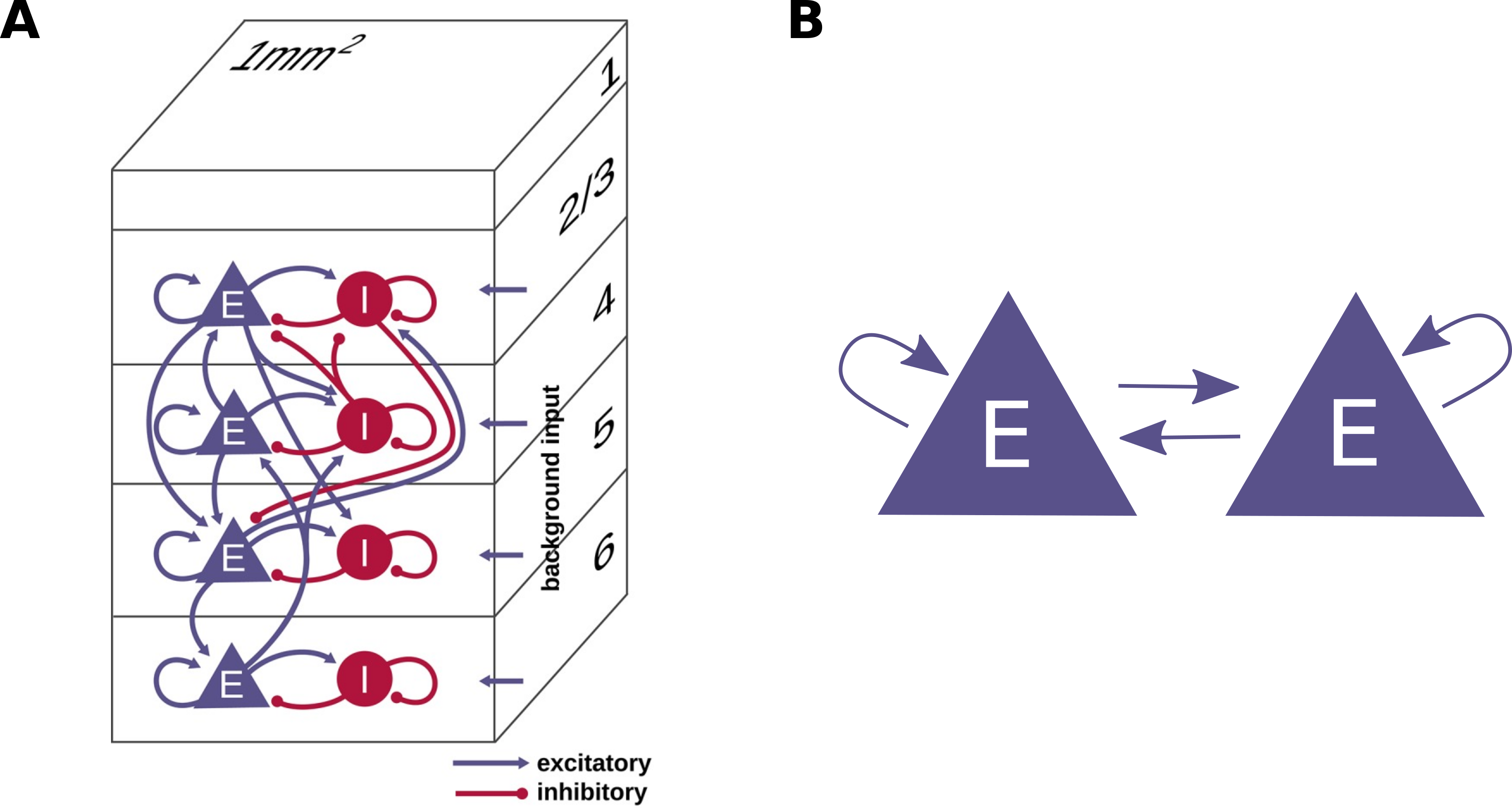}
\caption{
Schematic representation of the networks used in this work. \textbf{(A)}~Diagram of the cortical microcircuit model; reproduced from \cite{vanAlbada2018}. \textbf{(B)}~Scheme of the network of two populations of Izhikevich neurons.
\label{fig:simple_net}}
\end{figure}

The present work evaluates the performance of the proposed approach on two network models: a cortical microcircuit and a simple network model of two neuron populations. The models are depicted schematically in Figure \ref{fig:simple_net}.
The microcircuit model of Potjans and Diesmann \citep{Potjans_2014} represents a $1$\,mm$^2$ patch of early sensory cortex at the biological plausible density of neurons and synapses. The full-scale model comprises four cortical layers (L2/3, L4, L5, and L6) and consists of about $77,000$ current-based leaky-integrate-and-fire model neurons, which are organized into one excitatory and one inhibitory population per layer. These eight neuron populations are recurrently connected by about $300$ million synapses with exponentially decaying postsynaptic currents; the connection probabilities are derived from anatomical and electrophysiological measurements. 
The connection rule used is \texttt{fixed\_total\_number} with autapses and multapses allowed.
The dynamics of the membrane potentials and synaptic currents are integrated using the exact integration method proposed by Rotter and Diesmann \citep{Rotter_1999} and the membrane potential of the neurons of every population are initialized from a normal distribution with mean and standard deviation optimized from the neuron population as in \cite{vanAlbada2018}. This approach avoids transients at the beginning of the simulation. Signals originating from outside of the local circuitry, i.e., from other cortical areas and the thalamus, can be approximated with Poisson-distributed spike input or DC current input. Tables 1--4 of \cite{Dasbach2021} (see \textit{fixed total number} models) contain a detailed model description and report the values of the parameters. The model explains the experimentally observed cell-type and layer-specific firing statistics, and it has been used in the past both as a building block for larger models \cite[e.g.,][]{Schmidt2018} and as a benchmark for several validation studies \citep{vanAlbada2018,Knight2018,Rhodes2019,Knight2021_pygenn,Golosio2021,Kurth2022,Heittmann2022}.

The second model is designed for testing the scaling performance of the network construction by changing the number of neurons and the number of connections in the network across biologically relevant ranges for different connection rules (see Section \ref{sec:methods_dyn_conns}; autapses and multapses allowed).
The model consists of two equally sized neuron populations, which are recurrently connected to themselves and to each other in four \texttt{nestgpu.Connect()} calls.
The total number of neurons in the network is $N$ (i.e., $N/2$ per population) and the target total number of connections is $N\times K$ connections, where $K$ is the target number of connections per neuron.
Dependent on the connection rule used, the instantiated networks may exhibit small deviations from these target values:

\begin{itemize}
    \item \texttt{fixed\_total\_number}:\\
    The total number of connections used in each connect call is set to $\lfloor N\times K / 4 \rfloor$.
    \item \texttt{fixed\_indegree}:\\
    The in-degree used in each connect call is set to $\lfloor K /2 \rfloor$.
    \item \texttt{fixed\_outdegree}:\\
    The out-degree used in each connect call is set to $\lfloor K /2 \rfloor$.
\end{itemize}

The network uses Izhikevich neurons \citep{Izhikevich2003}, but note that the studied scaling behavior is independent of the neuron model; likewise of the neuron, connection and simulation parameters. Indeed, the only parameters that have an impact on this scaling experiment are the total number of neurons and the number of connections per neuron (i.e., $N$ and $K$).

\subsection{Hardware and software of performance evaluation}
\label{sec:method_hardware_software}
As a reference, we implement the proposed method for generating connections directly in GPU memory in the GPU version of the simulation code NEST. In the following, NEST GPU \textit{(onboard)} refers to the  new algorithm in which the connections are created directly in GPU memory, while NEST GPU \textit{(offboard)} indicates the previous algorithm which first generates the network in CPU memory and subsequently copies the network structure into the GPU as done in \cite{Golosio2021,Tiddia2022}. For a quantitative comparison to other established codes, we use the CPU version of NEST \citep{Gewaltig_07_11204} (version 3.3 \citep{NEST3.3}) and the GPU code generator GeNN \citep{Yavuz2016} (version 4.8.0\footnote{\url{https://github.com/genn-team/genn/releases/tag/4.8.0}}).

We evaluate the performance of the alternative codes on four systems equipped with NVIDIA GPUs of different generations and main application areas: two compute clusters, JUSUF \citep{VonStVieth2021} and JURECA-DC \citep{Thoernig2021}, both using CUDA version 11.3 and equipped with the data center GPUs V100 and A100, respectively, and two workstations with the consumer GPUs RTX 2080 Ti, with CUDA version 11.7 and RTX 4090 with CUDA version 11.4. The NEST GPU and GeNN simulations discussed in this work each employ a single GPU card, both because the novel network construction method developed for NEST GPU is limited to single-GPU simulations and also because all the simulation systems employed have enough GPU memory to simulate the models previously described using a single GPU card.
The CPU simulations use a single compute node of the HPC cluster JURECA-DC and exploit its 128 cores by 8 MPI processes each running 16 threads. Table \ref{tab:hardware_conf} shows the specifications of these three systems.

\begin{table}[H] 
\caption{Hardware configuration of the different systems used to measure the performance of the simulators. Cluster information is given on a per node basis.\label{tab:hardware_conf}}
\newcolumntype{C}{>{\centering\arraybackslash}X}
\begin{tabularx}{\textwidth}{CCC}
\toprule
\textbf{System}	& \textbf{CPU}	& \textbf{GPU}\\
\midrule
JUSUF cluster	& 2× AMD EPYC 7742, 2× 64 cores, 2.25 GHz	& NVIDIA V100\textsuperscript{1}, 1530 MHz, 16 GB HBM2e, 5120 CUDA cores \\
\midrule
JURECA-DC cluster	& 2× AMD EPYC 7742, 2× 64 cores, 2.25 GHz	& NVIDIA A100\textsuperscript{2}, 1410 MHz, 40 GB HBM2e, 6912 CUDA cores \\
\midrule
Workstation 1	& Intel Core  i9-9900K, 8 cores, 3.60 GHz	& NVIDIA RTX 2080 Ti\textsuperscript{3}, 1545 MHz, 11 GB GDDR6, 4352 CUDA cores  \\
\midrule
Workstation 2	& Intel Core  i9-10940X, 14 cores, 3.30 GHz	& NVIDIA RTX 4090\textsuperscript{4}, 2520 MHz, 24 GB GDDR6X, 16384 CUDA cores  \\
\bottomrule
\end{tabularx}
\noindent{\footnotesize{\textsuperscript{1} Volta architecture: \url{https://developer.nvidia.com/blog/inside-volta}}}\\
\noindent{\footnotesize{\textsuperscript{2} Ampere architecture: \url{https://developer.nvidia.com/blog/nvidia-ampere-architecture-in-depth}}}\\
\noindent{\footnotesize{\textsuperscript{3} Turing architecture: \url{https://developer.nvidia.com/blog/nvidia-turing-architecture-in-depth}}}\\
\noindent{\footnotesize{\textsuperscript{4} Ada Lovelace architecture: \url{https://www.nvidia.com/en-us/geforce/ada-lovelace-architecture}}}
\end{table}

For the network models, their specific implementations were taken from the original source for each simulator in the case of the cortical microcircuit model.
In particular, both NEST\footnote{\url{https://github.com/nest/nest-simulator}} and NEST GPU\footnote{\url{https://github.com/nest/nest-gpu}} provide an example implementation of the cortical microcircuit model inside their respective source code repositories.
Additionally, GeNN provides their own implementation of the microcircuit along with the data used for their PyGeNN publication \citep{Knight2021_pygenn} in the corresponding publicly available GitHub repository\footnote{\url{https://github.com/BrainsOnBoard/pygenn_paper}}.
Furthermore, to correctly compare the performance of the simulation, we adapt the existing scripts so that the overall behavior remains the same. In particular we disabled spike recordings and we enabled optimized initialization of membrane potentials as in \cite{vanAlbada2018}.\\
The second model presented in Section \ref{sec:methods_models} is implemented for NEST GPU \textit{(onboard)} and different connection rules can be chosen for the simulation.
%devised by the authors in order to evaluate the scaling performance of the new network construction method using a variable number of neurons and connections per neuron.
See the Data Availability Statement for further details on how to access the specific model versions used for this publication.

\subsection{Simulation phases}
\label{sec:methods_phases}
On the coarse level, we divide a network simulation into two successive phases: \textit{network construction} and \textit{simulation}.
The \textit{network construction} phase encompasses all steps until the actual simulation loop starts.
To assess different contributions to the network construction, we further divide this phase into stages. The consecutively executed stages in the NEST implementations (both CPU and GPU versions) follow the same pattern:
\begin{enumerate}
    \item \textit{initialization} is a setup phase in the Python script for preparing both model and simulator by importing modules, instantiating a class, or setting parameters, etc.:
    \begin{itemize}
        \item[] \texttt{import nestgpu}
    \end{itemize}
    \item \textit{node creation} instantiates all the neurons and devices of the model:
    \begin{itemize}
        \item[] \texttt{nestgpu.Create()}
    \end{itemize}
    \item \textit{node connection} instantiates the connections among network nodes:
        \begin{itemize}
        \item[] \texttt{nestgpu.Connect()}
    \end{itemize}
    \item \textit{calibration} is a preparation phase which orders the connections and initializes data structures for the spike buffers and the spike arrays just before the state propagation begins. In the CPU code, the pre-synaptic connection infrastructure is set up here. This stage can be triggered by simulating just one time step $h$.
        \begin{itemize}
        \item[] \texttt{nestgpu.Simulate(h)}
        \end{itemize}
        Previously, the calibration phase of NEST GPU was used to finish moving data to the GPU memory and instantiate additional data structures like the spike buffer (cf. Section \ref{sec:methods_spike_buffer}). Now, as no data transfer is needed and connection sorting is done instead (cf. Section \ref{sec:methods_orga_conns}), the calibration phase is now conceptually closer to the operations carried out in the CPU version of NEST \citep{Jordan2018}.\\
\end{enumerate}

In GeNN, the network construction is decomposed as follows:
\begin{enumerate}
    \item \textit{model definition} defines neurons and devices and synapses of the network model:
    \begin{itemize}
        \item[] \texttt{from pygenn import genn\_model}
        \item[] \texttt{model = genn\_model.GeNNModel()}
        \item[] \texttt{model.add\_neuron\_population()}
    \end{itemize}
    \item \textit{building} generates and compiles the simulation code:
        \begin{itemize}
        \item[] \texttt{model.build()}
    \end{itemize}
    \item \textit{loading} allocates memory and instantiates the network on the GPU:
    \begin{itemize}
        \item[] \texttt{model.load()}
    \end{itemize}
\end{enumerate}

Timers at the level of the Python interface assess the performance of the three different simulation engines. This has the advantage of being: agnostic to the implementation details of each stage at the kernel level, including any overhead of data conversion required by the C++ API, and close to the actual time perceived by the user.

\subsection{Validation of the proposed network construction method}
\label{sec:methods_validation}
The generation of random numbers for the probabilistic connection rules differs between the previous and the novel approach for network construction in NEST GPU.
This means that the connectivity resulting from the same rule with the same parameters is not identical, but only matches on a statistical level.
It is therefore necessary to determine that the network dynamics is qualitatively preserved.

Using the cortical microcircuit model, we validate the novel method against the previous one by means of a statistical analysis of the simulated spiking activity data.
To this end, we apply a similar validation procedure to that proposed in \cite{Golosio2021,Tiddia2022}, where the GPU version of NEST was compared to the CPU version as a reference.
We follow the example of \cite{vanAlbada2018, Knight2018, Dasbach2021, Heittmann2022} and compute for each of the eight neuron populations three statistical distributions to characterize the spiking activity:
\begin{itemize}
    \item time-averaged firing rate for each neuron
    \item coefficient of variation of inter-spike-intervals (CV ISI)
    \item pairwise Pearson correlation of the spike trains obtained from a subset of 200 neurons for each population.
\end{itemize}
These distributions are then compared for the two different approaches for network construction, as detailed in Appendix \ref{sec:appendix_validation}.

\section{Results}
\label{sec:results}

This section evaluates the performance of the proposed method for generating connections directly in GPU memory using the reference implementation NEST GPU \textit{(onboard)}.
For the cortical microcircuit model, we compare the network construction time and the real-time factor of the simulations obtained with the novel method to NEST GPU \textit{(offboard)} (i.e., the simulator version employing the previous algorithm of instantiating the connections first on the CPU), the CPU version of the simulator NEST \citep{Gewaltig_07_11204} and the code-generation based simulator GeNN \citep{Yavuz2016}.
\begin{figure}[t]
\centering
\includegraphics[width=0.95\textwidth]{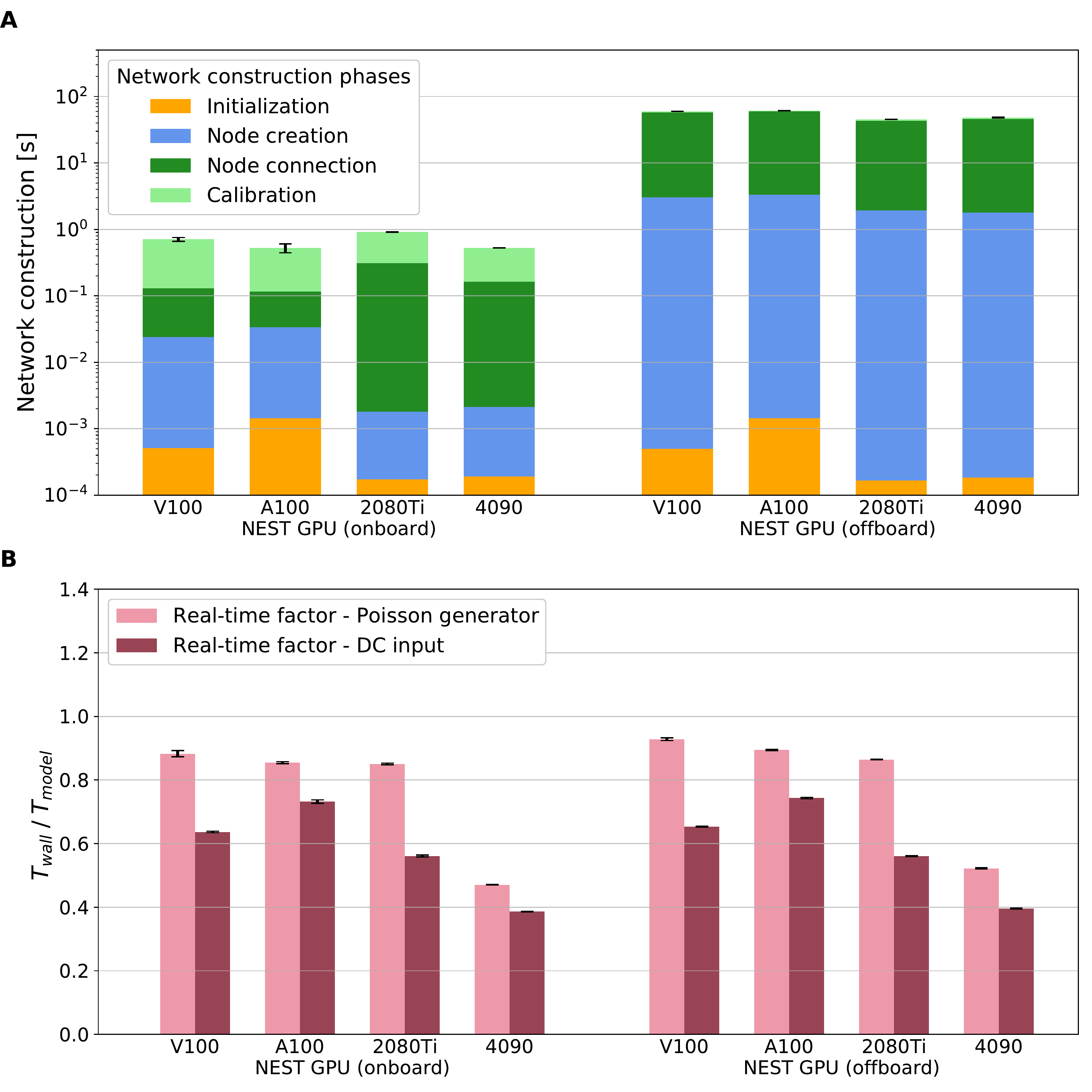}
\caption{Comparison of network construction phase and simulation of the network dynamics for the two versions of NEST GPU on the cortical microcircuit model. \textbf{(A)}~Performance comparison of the network construction phase using different hardware configurations. \textbf{(B)}~Real-time factor, defined as $T_{\text{wall}}/T_{\text{model}}$. The biological model time we use to compute the real-time factor is $T_{\text{model}}=10$\,s. The external drive is provided by Poisson spike generators (left bars, pink) or DC input (right bars, dark red).
Error bars show the standard deviation of the simulation phase over ten simulations using different random seeds.
\label{fig:microcircuit_nestgpu_comparison}}
\end{figure}
With the two-population network model, we assess the network construction time upon scaling the number of neurons and the number of connections per neuron.
Refer to Section \ref{sec:methods_models} for details on the network models.

\subsection{Cortical microcircuit model}
\label{sec:results_microcircuit}

Figure \ref{fig:microcircuit_nestgpu_comparison} directly compares the two approaches for network construction implemented in NEST GPU, i.e., \textit{onboard} and \textit{offboard}, in terms of the network construction time (panel A) and the real-time factor obtained by a simulation of the network dynamics (panel B).
Panel A shows that the novel method for network construction enables a speed-up by two orders of magnitude with respect to the previous network construction algorithm. 
While the \textit{offboard} method \cite[used in][]{Golosio2021,Tiddia2022} already optimizes the network construction on the CPU via multi-threading parallelization with the OpenMP library, the overhead of transferring the network from CPU to GPU becomes obsolete with the proposed approach to generate the connections directly on the GPU.
Moreover, the \textit{onboard} version shows lower network construction times across all hardware configurations without compromising the simulation times (panel B).
An additional detail to take note of, with the novel algorithm the calibration phase is now by far the longest compared to the node creation and node connection (3--5 times longer depending on the hardware used). However, this is only due to the fact that both creation and connection phases are now only used to instantiate data structures in GPU memory whereas the calibration phase takes charge of the connection sorting as described in Section \ref{sec:methods_orga_conns}.
Both versions have real-time factors of less than one second (sub-realtime simulation), thus showing also an improvement on the simulation time compared to the results of \cite{Golosio2021} obtained with the prototype NeuronGPU library.
Additionally, in some cases it is possible to see a small improvement when simulating using the novel network construction approach due to some code optimization related to the simulation phase.
While the network construction times are independent of the choice of external drive, the DC input as expected leads to faster simulations of the network dynamics compared to the Poisson generators.
Comparing the different hardware configuration, the smallest real-time factor obtained with NEST GPU \textit{(onboard)} is achieved with DC input on the latest consumer GPU RTX 4090, $0.386(0.001)$ (mean (standard deviation)). The respective result for Poisson input is $0.4707(0.0008)$.
For completeness, we also measure the real-time factor of NEST and GeNN simulations using the same framework used for Figure \ref{fig:microcircuit_nestgpu_comparison}B. These results are shown in Tables \ref{tab:performance_metrics_NEST*}, \ref{tab:performance_metrics_NEST*_DC}, and \ref{tab:performance_metrics_GeNN} and depicted in Appendix \ref{sec:appendix_microcircuit}.

\begin{figure}[H]
\centering
\includegraphics[width=0.95\textwidth]{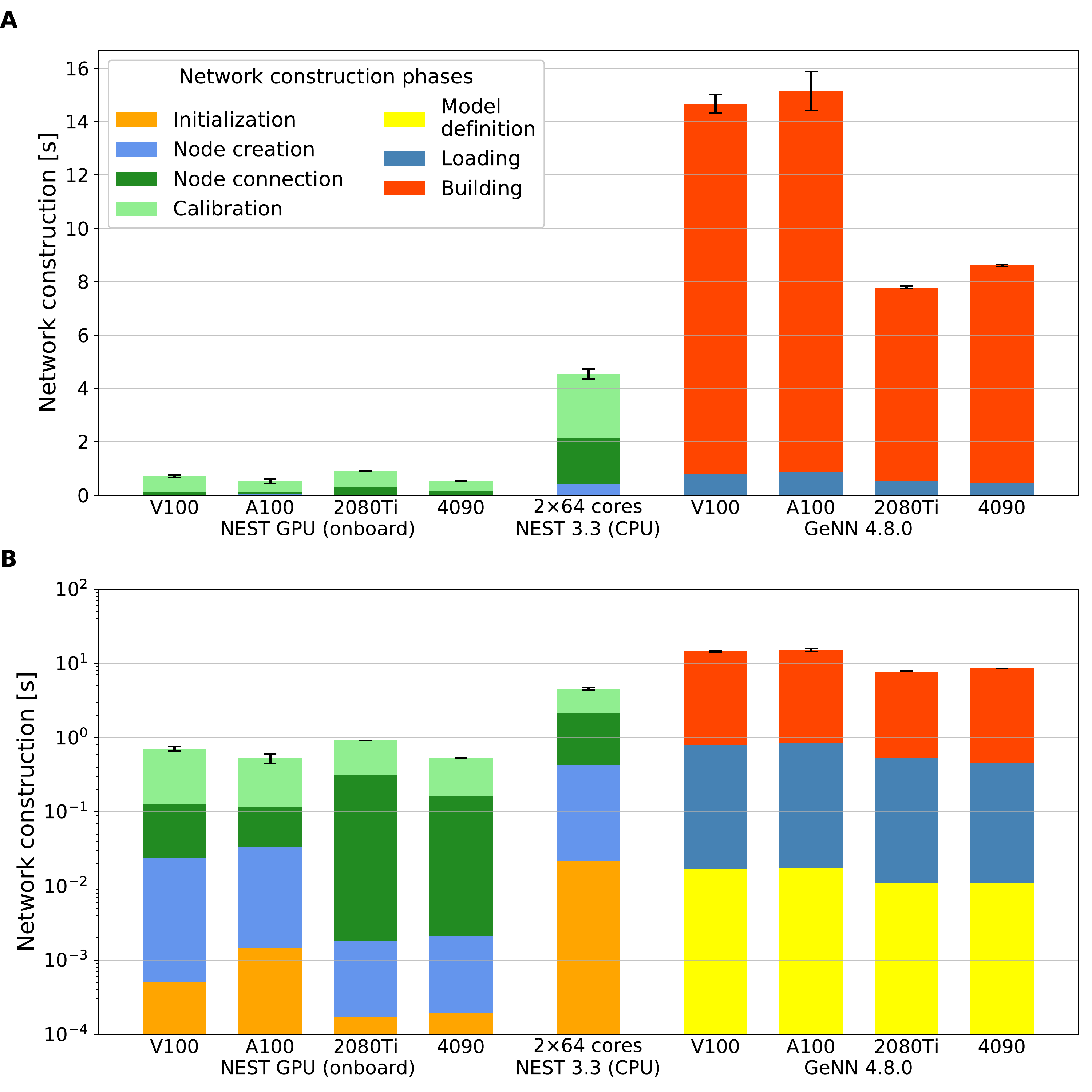}
\caption{
Performance comparison of the network construction phase for different simulators and hardware configurations on the cortical microcircuit model. Data for NEST GPU \textit{(onboard)} is the same as in Figure \ref{fig:microcircuit_nestgpu_comparison}.
\textbf{(A)}~Network construction time of the model in linear scale for different simulators and hardware configurations.
\textbf{(B)}~as in \textbf{(A)}~but with logarithmic y-axis scale. \label{fig:microcircuit_netw_construct}
In both panels, the \textit{building} phase of GeNN is placed on top of the bar, breaking with the otherwise chronological order, because this phase is not always required and at the same time, this display makes the shorter \textit{loading} phase visible in the plot with the logarithmic y-axis.
Error bars show the standard deviation of the overall network construction phase over ten simulations using different random seeds.
}
\end{figure}

Figure \ref{fig:microcircuit_netw_construct} compares the network construction times of the full-scale cortical microcircuit model obtained using NEST GPU \textit{(onboard)}, NEST 3.3 and GeNN 4.8.0 on different hardware configurations; for details on the hardware and software configurations see Section \ref{sec:method_hardware_software}. The settings are the same as in Figure \ref{fig:microcircuit_nestgpu_comparison}.
While panel A resolves the contributions of the different stages (defined in Section \ref{sec:methods_phases}) using a linear y-axis, panel B shows the same data with logarithmic y-axis to facilitate a comparison of the absolute numbers.
As mentioned in Section \ref{sec:methods_models}, the external input to the cortical microcircuit implementations in both the CPU and GPU version of NEST can be provided through either generators of Poisson signals or DC input. We run simulations comparing both approaches, however, Figure \ref{fig:microcircuit_netw_construct} only shows results for the case of Poisson generators because the network construction times with DC input are similar. GeNN in contrast mimics incoming Poisson spike trains through a current directly applied to each neuron.
Both NEST GPU \textit{(onboard)} and GeNN (without the \textit{building} phase) achieve fast network construction times of less than a second. The fastest overall network construction takes $0.499(0.10)$\,s as measured with NEST GPU \textit{(onboard)} using DC input on the A100 GPU, the data center GPU of the latest architecture tested. The time measured using the RTX 4090 is also compatible with the A100 result; the measured times with the V100 and the consumer GPU RTX 2080 Ti are also close.
Tables \ref{tab:performance_metrics_NEST*}, \ref{tab:performance_metrics_NEST*_DC}, and \ref{tab:performance_metrics_GeNN} provide the measured values for reference.

\begin{table}[H]
\caption{Performance metrics of NEST and NEST GPU when using Poisson spike generators to drive external stimulation to the neurons of the model. All times are in seconds with notation (mean (standard deviation)). Simulation time is calculated for a simulation of $10$\,s of biological time.\label{tab:performance_metrics_NEST*}}
\begin{adjustwidth}{-\extralength}{0cm}
\newcolumntype{C}{>{\centering\arraybackslash}X}
\begin{tabularx}{\fulllength}{C||C|C|C|C|C|C|C|C|C}
\toprule
\multirow{2}{*}{\textbf{Metrics}}	& \multicolumn{4}{c|}{\textbf{NEST GPU (onboard)}}	& \multicolumn{4}{c|}{\textbf{NEST GPU (offboard)}}	&  \textbf{NEST 3.3 (CPU)}\\
\cmidrule(l){2-10}
 & V100 & A100 & 2080Ti & 4090 & V100 & A100 & 2080Ti & 4090 & $2\times 64$ cores \\
\midrule % ------------------------------------
Initializa-
\newline tion &
\begin{math}
5.08(0.15)
\newline
\cdot10^{-4}
\end{math} & % V100
\begin{math}
1.44(0.15)
\newline
\cdot10^{-3}
\end{math} & % A100
\begin{math}
1.71(0.09)
\newline
\cdot10^{-4}
\end{math} & % 2080Ti
\begin{math}
1.91(0.04)
\newline
\cdot10^{-4}
\end{math} & % 4090
\begin{math}
4.99(0.08)
\newline
\cdot10^{-4}
\end{math} & % V100
\begin{math}
1.44(0.15)
\newline
\cdot10^{-3}
\end{math} & % A100
\begin{math}
1.66(0.12)
\newline
\cdot10^{-4}
\end{math} & % 2080Ti
\begin{math}
1.84(0.05)
\newline
\cdot10^{-4}
\end{math} & % 4090
\begin{math}
0.02(0.01)
\end{math} \\ % 1 node
\midrule % ------------------------------------
Node creation &
\begin{math}
0.02(0.004)
\end{math} & % V100
\begin{math}
0.03(0.007)
\end{math} & % A100
\begin{math}
1.63(0.09)
\newline
\cdot10^{-3}
\end{math} & % 2080Ti
\begin{math}
1.94(0.02)
\newline
\cdot10^{-3}
\end{math} & % 4090
\begin{math}
3.02(0.02)
\end{math} & % V100
\begin{math}
3.32(0.05)
\end{math} & % A100
\begin{math}
1.93(0.04)
\end{math} &  % 2080Ti
\begin{math}
1.781
\newline
(0.018)
\end{math} &  % 4090
\begin{math}
0.39(0.02)
\end{math} \\ % 1 node
\midrule % ------------------------------------
Node connection &
\begin{math}
0.105
\newline
(0.0003)
\end{math} & % V100
\begin{math}
0.08(0.002)
\end{math} & % A100
\begin{math}
0.308
\newline
(0.009)
\end{math} & % 2080Ti
\begin{math}
0.1600
\newline
(0.0005)
\end{math} & % 4090
\begin{math}
54.65(0.11)
\end{math} & % V100
\begin{math}
56.02(0.27)
\end{math} & % A100
\begin{math}
41.16(0.28)
\end{math} & % 2080Ti
\begin{math}
44.2(0.7)
\end{math} & % 4090
\begin{math}
1.72(0.17)
\end{math} \\ % 1 node
\midrule % ------------------------------------
Calibration &
\begin{math}
0.57
\newline(0.001)
\end{math} & % V100
\begin{math}
0.408
\newline
(0.005)
\end{math} & % A100
\begin{math}
0.602
\newline
(0.0006)
\end{math} & % 2080Ti
\begin{math}
0.3638
\newline
(0.0004)
\end{math} & % 4090
\begin{math}
1.99(0.01)
\end{math} & % V100
\begin{math}
2.06(0.01)
\end{math} & % A100
\begin{math}
2.202(0.01)
\end{math} & % 2080Ti
\begin{math}
2.183
\newline
(0.014)
\end{math} & % 4090
\begin{math}
2.39(0.01)
\end{math} \\ % 1 node
\midrule % ------------------------------------
\textbf{Network construction} &
\begin{math}
0.708
\newline
(0.001)
\end{math} & % V100
\begin{math}
\textbf{0.52(0.08)} 
\end{math} & % A100
\begin{math}
0.91(0.09)
\end{math} & % 2080Ti
\begin{math}
\textbf{0.5259} 
\newline
\textbf{(0.0008)}
\end{math} & % 4090
\begin{math}
59.67(0.13)
\end{math} & % V100
\begin{math}
61.41(0.27)
\end{math} & % A100
\begin{math}
45.29(0.32)
\end{math} & % 2080Ti
\begin{math}
48.2(0.7)
\end{math} & % 4090
\begin{math}
4.54(0.18)
\end{math} \\ % 1 node
\midrule % ------------------------------------
Simulation \newline ($10$\,s) &
\begin{math}
8.82(0.09)
\end{math} & % V100
\begin{math}
8.54(0.03)
\end{math} & % A100
\begin{math}
8.504(0.02)
\end{math} & % 2080Ti
\begin{math}
4.707\newline
(0.008)
\end{math} & % 4090
\begin{math}
9.28(0.04)
\end{math} & % V100
\begin{math}
8.94(0.02)
\end{math} & % A100
\begin{math}
8.64(0.01)
\end{math} & % 2080Ti
\begin{math}
5.219
\newline
(0.018)
\end{math} & % 4090
\begin{math}
12.66(0.08)
\end{math} \\ % 1 node
\bottomrule % ------------------------------------
\end{tabularx}
\end{adjustwidth}
\end{table}

\begin{table}[H] 
\caption{Performance metrics of NEST and NEST GPU when using DC input to drive external stimulation to the neurons of the model. All times are in seconds with notation (mean (standard deviation)). Simulation time is calculated for a simulation of $10$\,s of biological time. \label{tab:performance_metrics_NEST*_DC}}
\begin{adjustwidth}{-\extralength}{0cm}
\newcolumntype{C}{>{\centering\arraybackslash}X}
\begin{tabularx}{\fulllength}{C||C|C|C|C|C|C|C|C|C}
\toprule
\multirow{2}{*}{\textbf{Metrics}}	& \multicolumn{4}{c|}{\textbf{NEST GPU (onboard)}}	& \multicolumn{4}{c|}{\textbf{NEST GPU (offboard)}}	&  \textbf{NEST 3.3 (CPU)}\\
\cmidrule(l){2-10}
 & V100 & A100 & 2080Ti & 4090 & V100 & A100 & 2080Ti & 4090 & $2\times 64$ cores \\
\midrule % ------------------------------------
Initializa-
\newline tion &
\begin{math}
5.04(0.13)
\newline
\cdot10^{-4}
\end{math} & % V100
\begin{math}
1.44(0.08)
\newline
\cdot10^{-3}
\end{math} & % A100
\begin{math}
1.75(0.16)
\newline
\cdot10^{-4}
\end{math} & % 2080Ti
\begin{math}
1.97(0.09)
\newline
\cdot10^{-4}
\end{math} & % 4090
\begin{math}
5.1(0.4)
\newline
\cdot10^{-4}
\end{math} & % V100
\begin{math}
1.5(0.4)
\newline
\cdot10^{-3}
\end{math} & % A100
\begin{math}
1.62(0.04)
\newline
\cdot10^{-4}
\end{math} & % 2080Ti
\begin{math}
1.86(0.04)
\newline
\cdot10^{-4}
\end{math} & % 4090
\begin{math}
0.018
\newline
(0.003)
\end{math} \\ % 1 node
\midrule % ------------------------------------
Node creation &
\begin{math}
7.0(0.5)
\newline
\cdot10^{-3}
\end{math} & % V100
\begin{math}
6.6(0.3)
\newline
\cdot10^{-3}
\end{math} & % A100
\begin{math}
1.43(0.13)
\newline
\cdot10^{-3}
\end{math} & % 2080Ti
\begin{math}
1.64(0.04)
\newline
\cdot10^{-3}
\end{math} & % 4090
\begin{math}
3.01(0.02)
\end{math} & % V100
\begin{math}
3.28(0.03)
\end{math} & % A100
\begin{math}
1.91(0.02)
\end{math} &  % 2080Ti
\begin{math}
1.79(0.03)
\end{math} &  % 4090
\begin{math}
0.392
\newline
(0.003)
\end{math} \\ % 1 node
\midrule % ------------------------------------
Node connection &
\begin{math}
0.1028
\newline
(0.0004)
\end{math} & % V100
\begin{math}
0.0790
\newline
(0.0013)
\end{math} & % A100
\begin{math}
0.31(0.02)
\newline
(0.009)
\end{math} & % 2080Ti
\begin{math}
0.1538
\newline
(0.0005)
\end{math} & % 4090
\begin{math}
54.65(0.17)
\end{math} & % V100
\begin{math}
55.89(0.19)
\end{math} & % A100
\begin{math}
40.8(0.5)
\end{math} & % 2080Ti
\begin{math}
44.2(0.7)
\end{math} & % 4090
\begin{math}
1.53(0.07)
\end{math} \\ % 1 node
\midrule % ------------------------------------
Calibration &
\begin{math}
0.5785
\newline
(0.0013)
\end{math} & % V100
\begin{math}
0.412
\newline
(0.008)
\end{math} & % A100
\begin{math}
0.6011
\newline
(0.0006)
\end{math} & % 2080Ti
\begin{math}
0.3632
\newline
(0.0003)
\end{math} & % 4090
\begin{math}
1.993
\newline
(0.012)
\end{math} & % V100
\begin{math}
2.059
\newline
(0.016)
\end{math} & % A100
\begin{math}
2.194
\newline
(0.015)
\end{math} & % 2080Ti
\begin{math}
2.181
\newline
(0.015)
\end{math} & % 4090
\begin{math}
2.352
\newline
(0.005)
\end{math} \\ % 1 node
\midrule % ------------------------------------
\textbf{Network construction} &
\begin{math}
0.6888
\newline
(0.0018)
\end{math} & % V100
\begin{math}
\textbf{0.499(0.10)}
\end{math} & % A100
\begin{math}
0.91(0.02)
\end{math} & % 2080Ti
\begin{math}
\textbf{0.5189} 
\newline
\textbf{(0.0005)}
\end{math} & % 4090
\begin{math}
59.65(0.19)
\end{math} & % V100
\begin{math}
61.23(0.19)
\end{math} & % A100
\begin{math}
44.9(0.5)
\end{math} & % 2080Ti
\begin{math}
48.1(0.7)
\end{math} & % 4090
\begin{math}
4.30(0.07)
\end{math} \\ % 1 node
\midrule % ------------------------------------
Simulation \newline ($10$\,s) &
\begin{math}
6.36(0.02)
\end{math} & % V100
\begin{math}
7.32(0.05)
\end{math} & % A100
\begin{math}
5.61(0.03)
\end{math} & % 2080Ti
\begin{math}
3.86(0.01)
\end{math} & % 4090
\begin{math}
6.530\newline
(0.012)
\end{math} & % V100
\begin{math}
7.43(0.02)
\end{math} & % A100
\begin{math}
5.604\newline
(0.016)
\end{math} & % 2080Ti
\begin{math}
3.953\newline
(0.013)
\end{math} & % 4090
\begin{math}
7.77(0.15)
\end{math} \\ % 1 node
\bottomrule % ------------------------------------
\end{tabularx}
\end{adjustwidth}
\end{table}

\begin{table}[H] 
\caption{Performance metrics of GeNN. All times are in seconds with notation (mean (standard deviation)). Simulation time is calculated for a simulation of $10$\,s of biological time.\label{tab:performance_metrics_GeNN}}
\newcolumntype{C}{>{\centering\arraybackslash}X}
\begin{tabularx}{\textwidth}{C||C|C|C|C}
\toprule
\multirow{2}{*}{\textbf{Metrics}}	& \multicolumn{4}{c}{\textbf{GeNN}}\\
\cmidrule(l){2-5}
 & V100 & A100 & 2080Ti & 4090\\
\midrule % ------------------------------------
Model definition &
\begin{math}
1.704(0.008)
\newline
\cdot10^{-2}
\end{math} & % V100
\begin{math}
1.75(0.01)
\newline
\cdot10^{-2}
\end{math} & % A100
\begin{math}
1.07(0.01)
\newline
\cdot10^{-2}
\end{math} & % 2080Ti
\begin{math}
1.094(0.007)
\newline
\cdot10^{-2}
\end{math} \\ % 4090
\midrule % ------------------------------------
Building &
\begin{math}
13.87(0.36)
\end{math} & % V100
\begin{math}
14.301(0.72)
\end{math} & % A100
\begin{math}
7.25(0.04)
\end{math} & % 2080Ti
\begin{math}
8.15(0.04)
\end{math} \\ % 4090
\midrule % ------------------------------------
Loading &
\begin{math}
0.77(0.02)
\end{math} & % V100
\begin{math}
0.85(0.006)
\end{math} & % A100
\begin{math}
0.51(0.01)
\end{math} & % 2080Ti
\begin{math}
0.445(0.015)
\end{math} \\ % 4090
\midrule
Network construction (no building) &
\begin{math}
0.79(0.02)
\end{math} & % V100
\begin{math}
0.85(0.006)
\end{math} & % A100
\begin{math}
0.52(0.01)
\end{math} & % 2080Ti
\begin{math}
0.456(0.015)
\end{math} \\ % 4090
\midrule
Network construction &
\begin{math}
14.67(0.35)
\end{math} & % V100
\begin{math}
15.15(0.72)
\end{math} & % A100
\begin{math}
7.78(0.04)
\end{math} & % 2080Ti
\begin{math}
8.61(0.04)
\end{math} \\ % 4090
\midrule
Simulation \newline ($10$\,s) &
\begin{math}
6.48(0.01)
\end{math} & % V100
\begin{math}
5.39(0.01)
\end{math} & % A100
\begin{math}
7.007(0.01)
\end{math} & % 2080Ti
\begin{math}
2.719(0.006)
\end{math} \\ % 4090
\bottomrule % ------------------------------------
\end{tabularx}
\end{table}

Hitherto we discussed the performance for both network construction and simulation of NEST GPU \textit{(onboard)} compared to NEST GPU \textit{(offboard)}, NEST and GeNN. Turning on the statistical analysis of the simulated activity, data shows a good agreement between NEST GPU \textit{(offboard)} and NEST GPU \textit{(onboard)} as well as between NEST GPU \textit{(onboard)} and NEST 3.3.
That means that differences between the compared simulator versions are of the same order as fluctuations due to the choice of different seeds in either of the codes (see Section \ref{sec:methods_validation} and Appendix \ref{sec:appendix_validation}).

\subsection{Two-population network}

\begin{figure}[H]
\centering
\includegraphics[width=\textwidth]{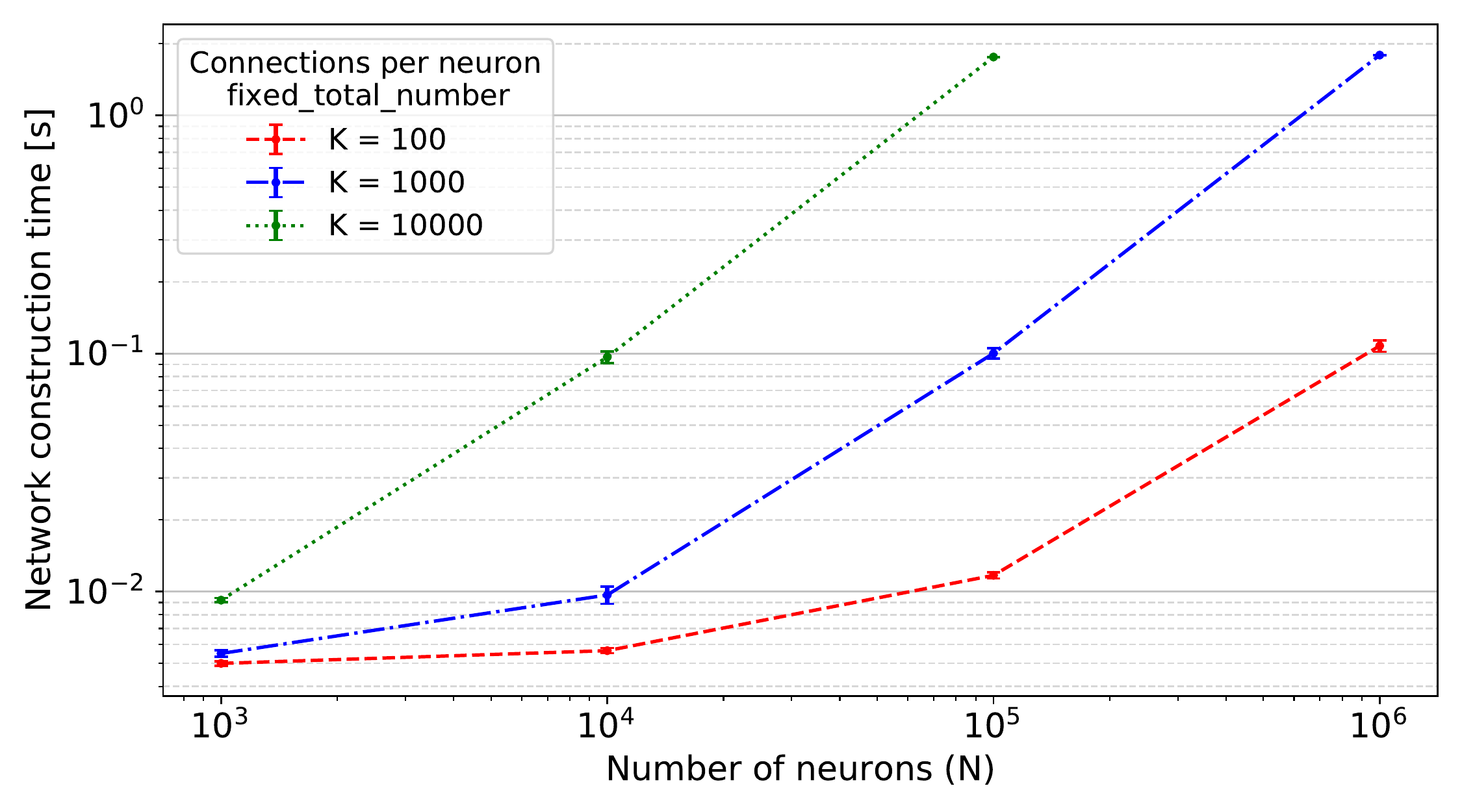}
\caption{Network construction time of the two-population network with $N$ neurons in total and $K$ connections per neuron using the \texttt{fixed\_total\_number} connection rule, i.e., the average amount of connections per neuron is $K$ and the total number of connections is $N\times K$. Error bars indicate the standard deviation of the performance across $10$ simulations using different seeds.
\label{fig:fixed_tot_num}}
\end{figure}

The two-population network described in Section \ref{sec:methods_models} is designed to evaluate the scaling performance of the proposed network construction method. To this end, we perform simulations on NEST GPU \textit{(onboard)} varying the number of neurons and the number of connections per neuron.
The scaling performance of NEST GPU \textit{(offboard)} has been evaluated on \cite{Golosio2021} for a balanced network model.
We opted for a total number of neurons in the network ($N$) ranging from $1,000$ to $1,000,000$ and a target number of connections per neuron ($K$) ranging from $100$ to $10,000$.

To enable the largest networks, benchmarks are performed on the JURECA-DC cluster, which is equipped with the GPUs with the largest GPU memory (i.e., the NVIDIA A100 with $40$\,GB) among the systems described in Table \ref{tab:hardware_conf}. Figure \ref{fig:fixed_tot_num} shows the network construction times using the \texttt{fixed\_total\_number} connection rule and ranging the number of neurons and connections per neuron. The performance obtained using the \texttt{fixed\_indegree} and \texttt{fixed\_outdegree} connection rules are totally compatible with the ones shown in this figure, and the respective plots are available in Appendix \ref{sec:appendix_izhikevich} for completeness.

As can be seen, the value of network construction time for the network with $10^6$ neurons and $10^4$ connections per neuron is not shown because of lack of GPU memory. Using an NVIDIA A100 GPU, we can thus say that this method enables the constructions of networks with up to an order of magnitude of $10^9$ connections.

\section{Discussion}
\label{sec:discussion}
It takes less than a second to generate the network of the cortical microcircuit model \citep{Potjans_2014} with the GPU version of NEST using our proposed dynamic approach for creating connections directly in GPU memory, on any GPU device tested. That is two orders of magnitude faster than the previous algorithm, which instantiates the connections first on the CPU and copies them from RAM to GPU memory just before the simulation starts (Figure \ref{fig:microcircuit_nestgpu_comparison}).
The reported network construction times are also shorter compared to the CPU version of NEST and the code generation framework GeNN (Figure \ref{fig:microcircuit_netw_construct}); if code generation and compilation are not required in GeNN, the results of NEST GPU and GeNN are compatible.
The time to simulate the network dynamics after network construction is not compromised by the novel approach.

The latest data center and consumer GPUs (i.e., A100 and RTX 4090, respectively) show the fastest network constructions as expected: approximately $0.5$\,s.
We observe the shortest simulation times on the RTX 4090 and attribute this result to the fact that the kernel design of NEST GPU particularly benefits from the high clock speeds of this device (cf. Section \ref{sec:method_hardware_software}).
Contrary to expectation, our simulations with DC input on the A100 are slower compared to the V100 although the former has higher clock speeds; an investigation of this observation is left for future work.

For models of the size of the cortical microcircuit, the novel approach renders the contribution of the network construction phase to the absolute wall-clock time negligible, even for short simulation durations. Further performance optimizations should preferentially rather target the simulation phase. Our result that GeNN currently simulates faster than NEST GPU indicates that there is room for improvement, which could possibly be exploited by further parallelization of the simulation kernel.

The evaluation of the scaling performance with the two-population network on the A100 shows that the network construction time is dominated by the total number of connections (i.e., $N\times K$, Figure \ref{fig:fixed_tot_num}) and mostly independent of the connection rule used.
The maximum network size that can be simulated depends on the GPU memory of the card employed for the simulation. Future generation GPU cards with more memory available will enable the construction of larger or denser networks of spiking neurons, and at the same time give reason to expect further performance improvements through novel architectures and the possibility of an even higher degree of parallelism. The novel approach is currently limited to simulations on a single GPU and future work is required to extend the algorithm to employing multiple GPUs as achieved with the previous algorithm \citep{Tiddia2022}.

Further improvements to the library may also expand upon the available connection rules and more flexible control via the user interface. At present, the pairwise Bernoulli connection routine \citep{Senk2022} is not available; this is because the onboard construction method requires a precise number of connections that must be allocated at once in order to not waste any GPU memory. The pairwise Bernoulli connection routine implies that this number is not known, hence additional heuristics would be required to optimize memory usage.
Autapses and multapses are currently always allowed in NEST GPU; therefore another useful addition would be the possibility to prohibit them (for example, using a flag as in the CPU version of NEST).

In conclusion, we propose a novel algorithm for network construction which dynamically creates the network exploiting the high degree of parallelism of GPU devices. It enables short network construction times comparable to code generation methods, advantageous flexibility of run-time instantiation of the network. This optimized method makes the contribution of network construction phase in network simulations marginal, even when simulating highly-connected large-scale networks. As discussed in \cite{Schmitt2023}, this is especially interesting for parameter scan applications, where a high volume of simulations needs to be tested and any additional contribution to the overall execution time of each test aggregates considerably and slows down the exploration process.

\vspace{6pt}

\section*{Author Contributions}
% all authors: B.G., J.V., G.T., E.P., J.St., V.F., P.S.P., A.M. and J.Se.
%
% Ideas; formulation or evolution of overarching research goals and aims.
Conceptualization, B.G., J.V., G.T., E.P., J.St., V.F., P.S.P., A.M. and J.Se.;
%
% Development or design of methodology; creation of models.
methodology, B.G., J.V., G.T., E.P., J.St., P.S.P., A.M. and J.Se;
%
% Programming, software development; designing computer programs; implementation of the computer code and supporting algorithms; testing of existing code components.
software, B.G., J.V., G.T., J.St. and J.Se.;
%
% formal analysis: Application of statistical, mathematical, computational, or other formal techniques to analyze or synthesize study data.
% investigation: Conducting a research and investigation process, specifically performing the experiments, or data/evidence collection.
% visualization: % Preparation, creation and/or presentation of the published work, specifically visualization/data presentation.
% validation: Verification, whether as a part of the activity or separate, of the overall replication/reproducibility of results/experiments and other research outputs.
% data curation: Management activities to annotate (produce metadata), scrub data and maintain research data (including software code, where it is necessary for interpreting the data itself) for initial use and later re-use.
investigation, formal analysis, visualization, validation and data curation, B.G., J.V., G.T. and J.Se.;
%
% resources:Provision of study materials, reagents, materials, patients, laboratory samples, animals, instrumentation, computing resources, or other analysis tools.
% funding acquisition: Acquisition of the financial support for the project leading to this publication.
% supervision: Oversight and leadership responsibility for the research activity planning and execution, including mentorship external to the core team.
resources, funding acquisition and supervision, B.G., P.S.P., A.M. and J.Se.;
% Preparation, creation and/or presentation of the published work, specifically writing the initial draft (including substantive translation).
writing---original draft preparation, B.G., J.V., G.T. and J.Se.;
%
%Preparation, creation and/or presentation of the published work by those from the original research group, specifically critical review, commentary or revision – including pre- or post-publication stages.
writing---review and editing, B.G., J.V., G.T., E.P., J.St., V.F., P.S.P., A.M. and J.Se.;
%
% Management and coordination responsibility for the research activity planning and execution.
project administration, B.G. and J.Se.
All authors have read and agreed to the published version of the manuscript.

\section*{Funding}
This project has received funding from
% HBP (SGA3) [to JS, PSP]
the European Union’s Horizon 2020 Framework Programme for Research and Innovation under Specific Grant Agreement No. 945539 (Human Brain Project SGA3),
% MetaMoSim [to JSe, JV]
the Initiative and Networking Fund of the Helmholtz Association in the framework of the Helmholtz Metadata Collaboration project call (ZT-I-PF-3-026),
and
% SMHB [to JSe, JSt, JV]
the Joint Lab “Supercomputing and Modeling for the Human Brain”, the Italian PNRR MUR project PE0000013-FAIR, funded by NextGenerationEU.

\section*{Data Availability Statement}
The code to reproduce all figures of this manuscript is publicly available at Zenodo:
% Use DOI for all versions
\url{https://doi.org/10.5281/zenodo.7744238}.
The versions of NEST GPU employed in this work are available on GitHub (\url{https://github.com/nest/nest-gpu}) via the git tags \texttt{nest-gpu\_onboard} and \texttt{nest-gpu\_offboard}.

\section*{Acknowledgements}
The authors gratefully acknowledge the computing time granted by the JARA Vergabegremium and provided on the JARA Partition part of the supercomputer JURECA at Forschungszentrum Jülich (computation grant JINB33), and the use of Fenix Infrastructure resources, which are partially funded from the European Union's Horizon 2020 research and innovation programme through the ICEI project under the Grant Agreement No. 800858.
The authors further thank the INFN APE Parallel/Distributed Computing laboratory and the the INM-6/IAS-6.
Part of the work was performed while Gianmarco Tiddia enjoyed a scientific stay at INM-6/IAS-6 in the period 21th of September 2022 to 28th of March 2023.
The authors would also like to thank Markus Diesmann for detailed comments on the manuscript and Andrea Bosin, Fabrizio Muredda and Giovanni Serra for their support in the use of the RTX 4090 GPU.

\section*{Conflicts of interest}
The authors declare no conflict of interest.
The sponsors had no role in the design, execution, interpretation, or writing of the study.

\section{Bibliography}
\bibliography{bibliography.bib}

\newpage
\appendix
\section[\appendixname~\thesection]{Block sorting}
\label{sec:appendix_block_sorting}
The following appendix describes the block sorting algorithm employed in the network construction phase of the simulation, and in particular when organizing connections among the nodes of the network.
\subsection{The COPASS (COnstrained PArtition of Sorted Subarrays) block-sort
  algorithm}

Given a real-number array $\textbf{A}$
divided in $k$ blocks (subarrays) $S_{i,j}$ of sizes $N_i$
\begin{align}
  \textbf{A} &= \left( S_{0,0}, \ldots, S_{0,N_0}, S_{1,0}, \ldots, S_{1,N_1},
  \ldots S_{k-1,0}, \ldots, S_{k-1,N_{k-1}} \right) \\
  S_{i,j} &\in \mathbb{R} \qquad i = 0, ..., k-1 \qquad j = 0, ..., N_i
\end{align}

  \begin{figure}[H]
%\begin{adjustwidth}{-\extralength}{0cm}
\centering
\includegraphics[width=\textwidth]{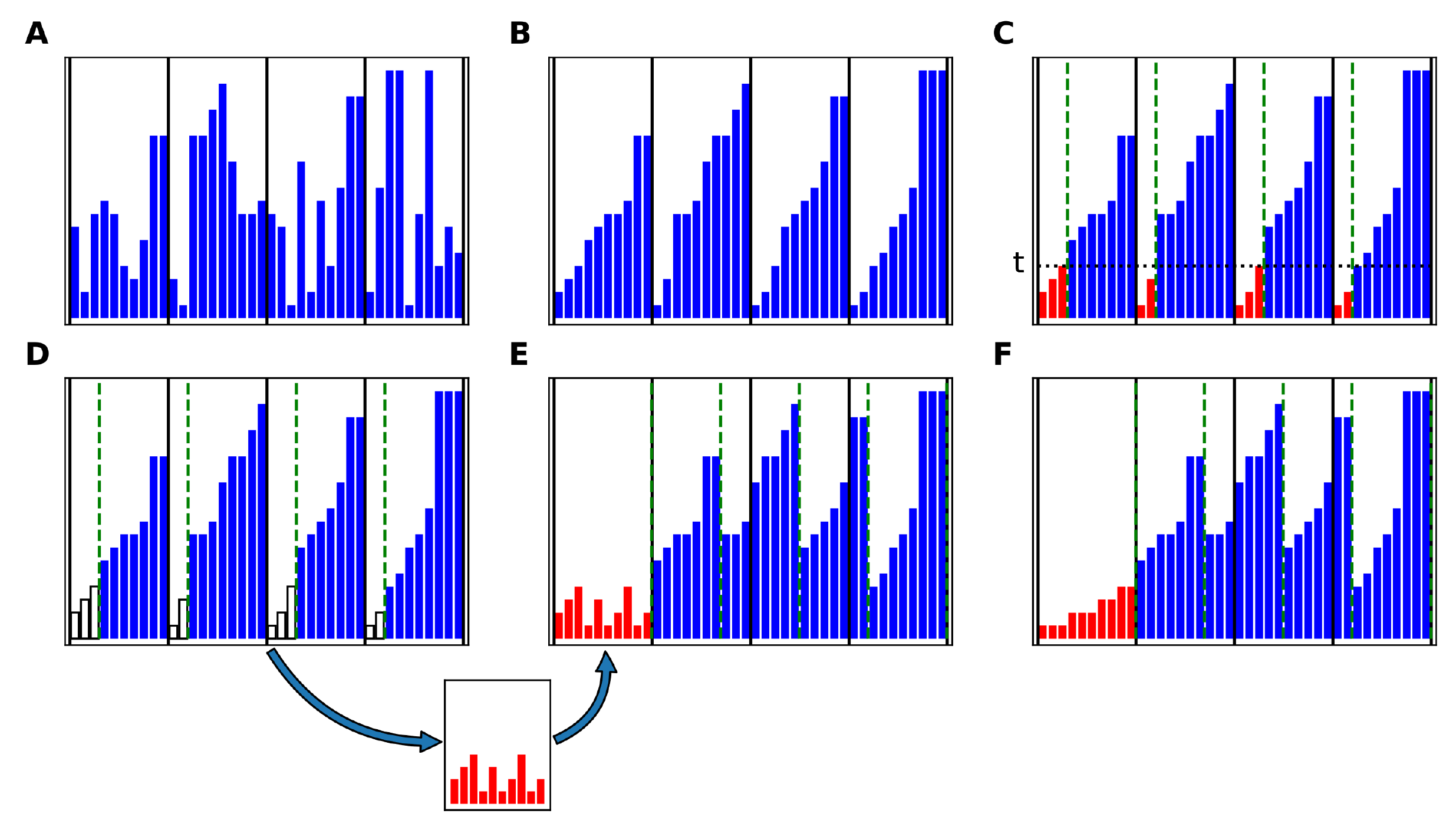}
%\end{adjustwidth}
\caption{
  The COPASS block-sort algorithm.
  \textbf{(A)} Unsorted array, divided in blocks (subarrays).
  Each element of the array is represented by a blue bar.
  The vertical solid lines represent the division in subarrays. 
  \textbf{(B)} Each subarray is sorted using the underlying sorting algorithm.
  \textbf{(B)} The subarrays are divided in two partitions each using a common
  threshold, $t$, in such a way that the total size of the left partitions
  (represented in red) is equal to the size of the first block.
  \textbf{(D)} The left partitions are copied to the auxiliary array.
  \textbf{(E)} The right partitions are shifted to the right,
  and the auxiliary array is copied to the first block.
  \textbf{(F)} The auxiliary array is sorted.
  The procedure from \textbf{(C)} to \textbf{(E)} is then repeated on the new
  subarrays, delimited by the green dashed lines, in order to extract and sort the second block, and so on until the last block.
  } \label{fig:COPASS}
\end{figure}
The aim of the COPASS block sort algorithm is to perform an in-place
sort of $\textbf{A}$ maintaining its block structure.
This algorithm relies on another algorithm for sorting each block.
It should be noted that the subarrays do not need to be stored in contiguous locations in memory.
The COPASS block-sort algorithm is illustrated in Figure \ref{fig:COPASS}.
The $k$ subarrays are sorted using the underlying sorting algorithm
(Figure \ref{fig:COPASS}B).
Each sorted subarrays is divided in two partitions  using the COPASS algorithm, described in the next sections, in such a way that
all the elements of the left partitions (represented in red)
are smaller than or equal to a proper common threshold, $t$,
while all the elements of the right partitions
are greater than or equal to $t$,
and the total number of elements of the left partitions is equal to
the size of the first block, $N_0$ (Figure \ref{fig:COPASS}C).
The elements of the left partitions are copied to the auxiliary array
(Figure \ref{fig:COPASS}D).
The right partitions are shifted to the right, leaving the first block
free, and the auxiliary array is copied to the first block
(Figure \ref{fig:COPASS}E).
The auxiliary array is sorted (Figure \ref{fig:COPASS}F).
The whole procedure is then repeated for extracting the second block,
using the logical subarrays delimited by the green dashed lines in
Fig. \ref{fig:COPASS}F, and so on until the last block is extracted.
The maximum size of the auxiliary array is equal to the size
of the largest block, i.e.,
\begin{align}
m_\text{max} = \textbf{max}_i \{ N_i \}
\end{align}
The auxiliary storage requirement of the COPASS block-sort algorithm
is the largest between the auxiliary storage requirement of the
underlying sorting algorithm for an array of size $m_\text{max}$ 
and the auxiliary array storage requirement.
This requirement can be reduced by dividing $\textbf{A}$ in a large number of small blocks.

\subsection{The COPASS partition algorithm}

Given a set of $k$ real-number arrays $S_{i,j}$
(here called {\it subarrays}) of sizes $N_i$
\begin{align}
  S_{i,j} \in \mathbb{R} \qquad i = 0, ..., k-1 \qquad j = 0, ..., N_i
\end{align}
each sorted in ascending order
\begin{align}
  S_{i,j} \le S_{i,l}  \qquad \text{for} \quad j \le l
\end{align}
and a positive integer
$m < \sum_{i,j} S_{i,j}$, %(total size of the left partitions),
the purpose of this algorithm is to find a threshold $t$
and $k$ non-negative integers $m_i$ %(left partition sizes)
such that
\begin{align}
  S_{i,j} &\le t   \qquad &\text{for} \quad & j < m_i \\
  S_{i,j} &\ge t   \qquad &\text{for} \quad & j \ge m_i \\    
  \sum_i {m_i} &= m
\end{align}
We will call {\it left partitions} the subarrays of size $m_i$
\begin{align}
  S_{i,j}  \qquad j=0, \ldots,  m_i - 1
\end{align}
and {\it right partitions} the complementary subarrays
\begin{align}
  S_{i,j} \qquad j=m_i , \ldots,  N_i - 1
\end{align}
The basic idea of the algorithm is to start from an initial interval
$[\ubar{t}_0 , \bar{t}_0]$
such that
$\ubar{t}_0 \le t \le  \bar{t}_0$
and to proceed iteratively, shrinking the interval
and ensuring that the condition
\begin{align}
\ubar{t}_s \le t \le  \bar{t}_s
\end{align}
is satisfied at each iteration index $s$, until
either $\ubar{t}_s$ or $\bar{t}_s$ is equal to $t$.
For this purpose, for each iteration index $s$ we define $\ubar{m}_{i,s}$
as the number of the elements of the subarray
$S_{i,j}$ that are smaller than or equal to $\ubar{t}_s$,
i.e., the cardinality of the set of integers $j$ such that $S_{i,j} \le \ubar{t}_s$
\begin{align} \label{Eq:ubar_m}
  \ubar{m}_{i,s} = \textbf{card} \{j: S_{i,j} \le \ubar{t}_s \}
\end{align}
and $\bar{m}_{i,s}$ as the number of elements
that are strictly smaller than $\bar{t}_s$
\begin{align} \label{Eq:bar_m}
  \bar{m}_{i,s} = \textbf{card} \{j: S_{i,j} < \bar{t}_s \}
\end{align}
Since the subarrays $S_{i,j}$ are sorted,
$\ubar{m}_{i,s}$ and $\bar{m}_{i,s}$
can be computed through a binary search algorithm.
In a parallel implementation, their values can be
evaluated for all $i = 0, \ldots, k-1$ by performing the binary searches
in parallel on the $k$ subarrays.
As an initial condition, we set
\begin{align}
\ubar{t}_0 &= \textbf{min}(S_{i,j}) - 1 \\
\bar{t}_0 &= \textbf{max}(S_{i,j}) + 1
\end{align}
From Eqs. \ref{Eq:ubar_m} and \ref{Eq:bar_m}, it follows that
\begin{align}
\ubar{m}_{i,0} &= 0 \\
\bar{m}_{i,0} &= N_i
\end{align}
and clearly 
\begin{align}
  \sum_i \ubar{m}_{i,0} <  m < \sum_i \bar{m}_{i,0}
\end{align}
We proceed iteratively to evaluate the values of 
$\ubar{t}_{s+1}$, $\bar{t}_{s+1}$, $\ubar{m}_{i,s+1}$ and $\bar{m}_{i,s+1}$
for the iteration index $s+1$ from their values at the previous iteration
index $s$.
Assume that the condition 
\begin{align} \label{Eq:m_range}
  \sum_i \ubar{m}_{i,s} <  m < \sum_i \bar{m}_{i,s}
\end{align}
is satisfied for the iteration index $s$.
The iterations are carried on only if 
$\bar{m}_{i,s} - \ubar{m}_{i,s} > 1$
for at least one index $i$, i.e. 
\begin{align} \label{Eq:m_diff_gt1}
  \exists i : \bar{m}_{i,s} - \ubar{m}_{i,s} > 1
\end{align}
If the latter condition is not met,  the iterations are concluded and a solution is found as described in
Section \ref{sec:sort_last_step_1}.
Otherwise, if Eq. \ref{Eq:m_diff_gt1} is satisfied, let
\begin{align}
l_s &= \textbf{arg max}_i \{ \bar{m}_{i,s} - \ubar{m}_{i,s} \} \\
\tilde{m}_s &= \lfloor \frac{ \ubar{m}_{l_s,s} + \bar{m}_{l_s,s} } {2} \rfloor \\
\tilde{t}_s &= S_{l_s, \tilde{m}_s}
\end{align}
where $\lfloor x \rfloor$ represents the integer part of $x$.
Since $\bar{m}_{l_s,s} - \ubar{m}_{l_s,s} > 1$,
clearly $\ubar{m}_{l_s,s} < \tilde{m}_s < \bar{m}_{l_s,s}$,
and from Eqs. \ref{Eq:ubar_m} and \ref{Eq:bar_m}
\begin{align}
  \ubar{t}_s < \tilde{t}_s <  \bar{t}_s
\end{align}
Let
\begin{align}
  \bar{\mu}_{i,s} &= \textbf{card} \{j: S_{i,j} \le \tilde{t}_s \} \label{eq:mubar} \\
  \ubar{\mu}_{i,s} &= \textbf{card} \{j: S_{i,j} < \tilde{t}_s \}
  \label{eq:muubar}
\end{align}
From the latter equations and from
Eqs. \ref{Eq:ubar_m} and \ref{Eq:bar_m} it follows that
\begin{align}
  \ubar{m}_{i,s} \le  \ubar{\mu}_{i,s} \le  \bar{\mu}_{i,s}  \le  \bar{m}_{i,s}
\end{align}
for all $i$, and thus
\begin{align}
  \sum_i \ubar{m}_{i,s} \le  \sum_i \ubar{\mu}_{i,s}
  \le  \sum_i \bar{\mu}_{i,s}  \le  \sum_i \bar{m}_{i,s}
\end{align}
%
%$m$ is somewhere in the whole interval.
Three cases are possible:
\begin{itemize}
  \item {\bf case 1}
    \begin{align} \label{eq:case1}
      \sum_i \ubar{\mu}_{i,s} \le  m \le \sum_i \bar{\mu}_{i,s}
    \end{align}
    in this case $t = \tilde{t}_s$.
    The iteration is concluded and the partition sizes $m_i$
    are computed using the procedure described in Section
    \ref{sec:sort_last_step_2}.
  \item {\bf case 2}
    \begin{align} \label{Eq:cond2}
      \sum_i \ubar{m}_{i,s} <  m < \sum_i \ubar{\mu}_{i,s}
    \end{align}
    In this case we set
    \begin{align}
      \ubar{m}_{i,s+1} &= \ubar{m}_{i,s} \qquad &\ubar{t}_{s+1} &= \ubar{t}_s \label{Eq:cond2_update1} \\
      \bar{m}_{i,s+1} &= \ubar{\mu}_{i,s} \qquad &\bar{t}_{s+1} &= \tilde{t}_s \label{Eq:cond2_update2}
    \end{align}
    and continue with the next iteration.
    Eqs. \ref{Eq:cond2}, \ref{Eq:cond2_update1} and \ref{Eq:cond2_update2}
    ensure that the condition of Eq. \ref{Eq:m_range}
    is satisfied for the next iteration index $s+1$.
  \item {\bf case 3}
    \begin{align} \label{Eq:cond3}
      \sum_i \bar{\mu}_{i,s}  <  m < \sum_i \bar{m}_{i,s}
    \end{align}
    In this case we set
    \begin{align}
      \ubar{m}_{i,s+1} &= \bar{\mu}_{i,s} \qquad &\ubar{t}_{s+1} &= \tilde{t}_s \label{Eq:cond3_update1} \\
      \bar{m}_{i,s+1} &= \bar{m}_{i,s} \qquad &\bar{t}_{s+1} &= \bar{t}_s
      \label{Eq:cond3_update2}
    \end{align}
    and continue with the next iteration.
    Eqs. \ref{Eq:cond3}, \ref{Eq:cond3_update1} and \ref{Eq:cond3_update2}
    ensure that the condition of Eq. \ref{Eq:m_range}
    is satisfied for the next iteration index $s+1$.

\end{itemize}

\subsection{The COPASS partition last step, case 1} \label{sec:sort_last_step_1}
This final step is carried out at the end of the iterations
when the following condition is met:
\begin{align}
  \bar{m}_{i,s} - \ubar{m}_{i,s} \le 1 \quad \forall i
\end{align}
Consider the set of the ordered pairs $(S_{i, \bar{m}_{i, s}}, i)$
such that $\bar{m}_{i, s}$ is equal to $\ubar{m}_{i, s} + 1$
\begin{align}
C = \{ (S_{i, \bar{m}_{i, s}}, i):
                     \bar{m}_{i, s} = \ubar{m}_{i, s} + 1 \}
\end{align}
We sort them in ascending order of $S_{i, \bar{m}_{i, s}}$ values 
\begin{align}
\tilde{C} = \textbf{sort} (C)
\end{align}
Let $d$ be the difference
\begin{align}\label{eq:m_diff}
d = m - \sum_i \ubar{m}_{i, s}
\end{align}
and $D$ the set of the first $d$ elements of $\tilde{C}$
\begin{align}
D = \{ \tilde{C}_0, \ldots, \tilde{C}_{d-1} \}
\end{align}
We set the left partition sizes as
\begin{align}
m_i &= \ubar{m}_{i, s} \qquad &\text{for} \quad
                       (S_{i, \bar{m}_{i, s}}, i) &\notin D \\
m_i &= \ubar{m}_i + 1 \qquad &\text{for} \quad
                       (S_{i, \bar{m}_{i, s}}, i) &\in D
\end{align}
From the latter equation, obviously the total size of the left partitions will
be
\begin{align}
\sum m_i = \sum \ubar{m}_{i, s} + d
\end{align}
and from Eq. \ref{eq:m_diff} it can be observed that this is equal to $m$,
as requested.
Furthermore, since $D$ is sorted,
the elements of the left partitions will be smaller than
or equal to those of the right partitions.

\subsection{The COPASS partition last step, case 2} \label{sec:sort_last_step_2}
This last step is taken when the condition of Eq. \ref{eq:case1} is met.
In this case, $t = \tilde{t}_s$, and from Eqs. \ref{eq:mubar},
\ref{eq:muubar} and \ref{eq:case1}, it follows that
\begin{align} \label{eq:S_eq_t}
S_{i, j} = t \qquad &\text{for} \quad
           \ubar{\mu}_{i,s} \le  j \le \bar{\mu}_{i,s}
\end{align}
Let $d$ be the difference
\begin{align}\label{eq:m_diff2}
d = m - \sum_i \ubar{\mu}_{i, s}
\end{align}
In order to find a solution for the left partition sizes, $m_i$,
we need to find $k$ integers, $d_i$, in the ranges
$[0, \bar{\mu}_{i,s} - \ubar{\mu}_{i,s}]$,
such that their sum is equal to $d$
\begin{align}
&\sum_i d_i = d \label{eq:sum_diff}\\
&\ubar{\mu}_{i,s} \le  d_i \le \bar{\mu}_{i,s}
\end{align}
and set
\begin{align}\label{eq:m_i}
m_i = \ubar{\mu}_{i,s} +  d_i
\end{align}
In fact, from Eqs. \ref{eq:m_diff2}, \ref{eq:sum_diff} and
\ref{eq:m_i} it follows that
\begin{align}
\sum_i m_i = m
\end{align}
as requested, while Eqs. \ref{eq:muubar} and \ref{eq:S_eq_t}
imply that $S_{i, j}$ is smaller than or equal to $t$ in the left partitions,
while it is larger than or equal to $t$ in the right partitions.

\begin{figure}[t]
\centering
\includegraphics[width=\textwidth]{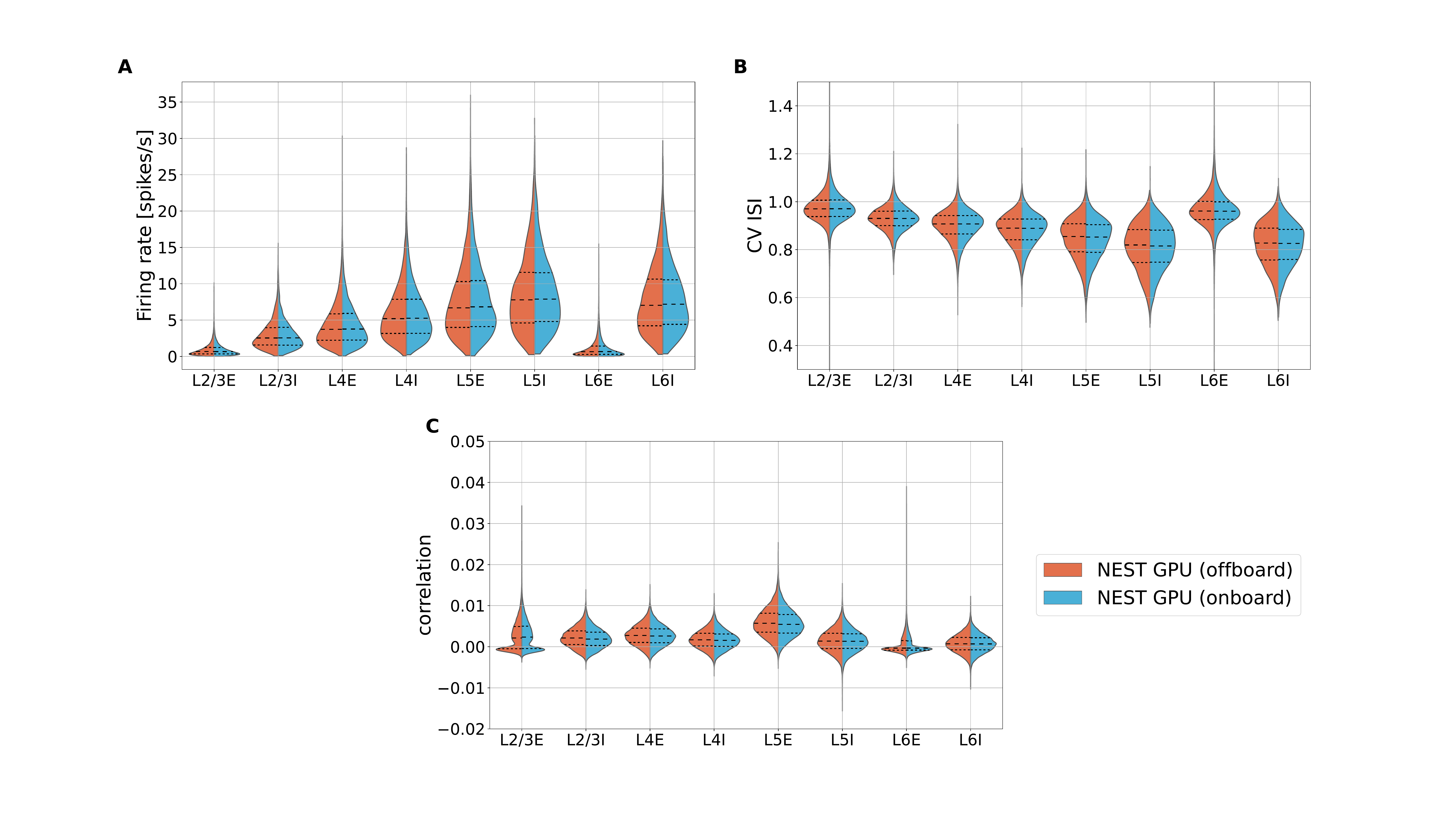}
\includegraphics[width=\textwidth]{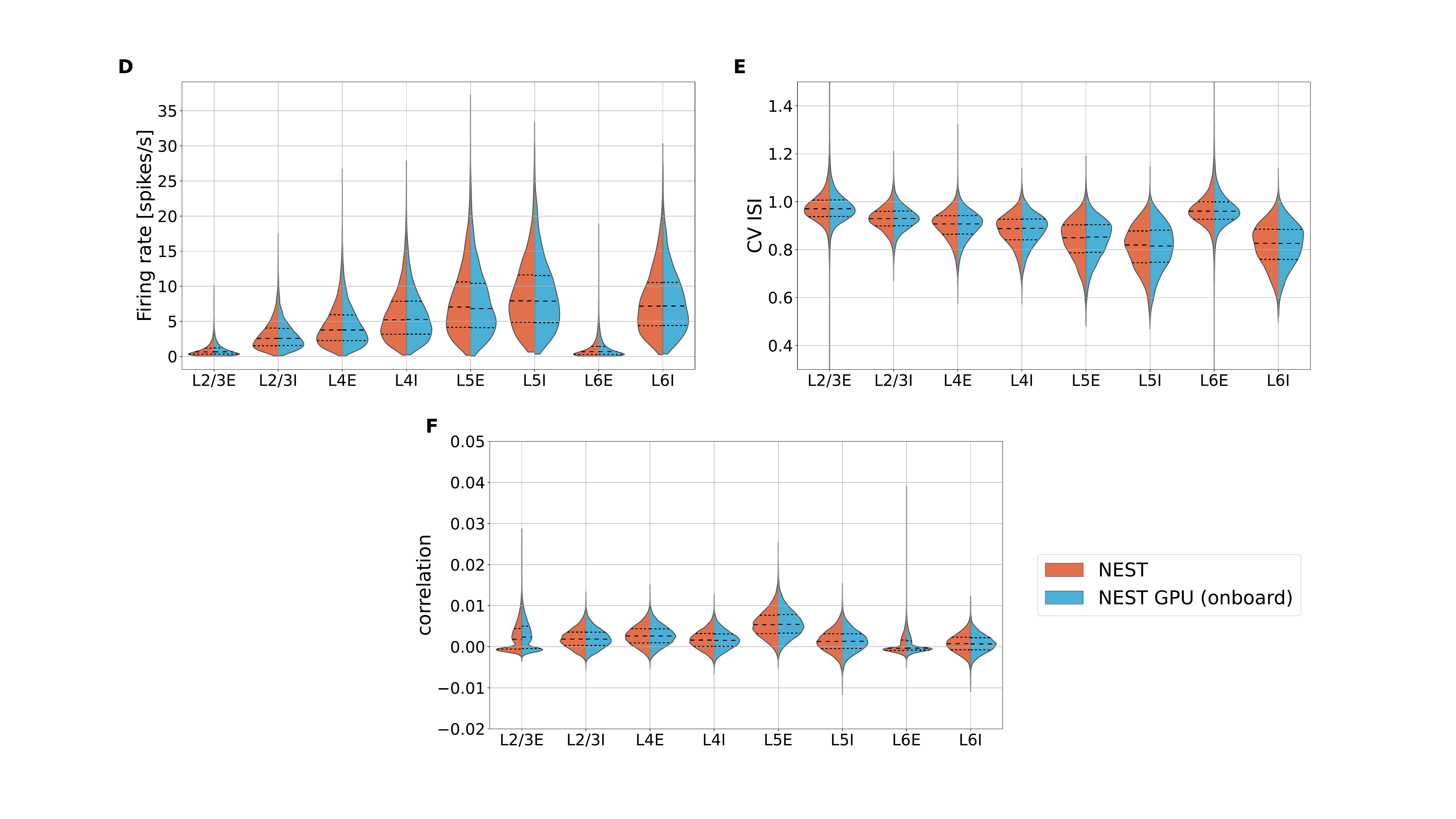}
\caption{
Violin plots of the distributions of firing rate \textbf{(A)}, CV ISI \textbf{(B)} and Pearson correlation \textbf{(C)} for a simulation for the populations of the cortical microcircuit model using NEST GPU with (sky blue distributions, right) or without (orange distributions, left) the new method for network construction. \textbf{(D, E, F)} Same as \textbf{(A, B, C)} but the orange distributions are obtained using NEST 3.3. Central dashed line represents the median of the distributions, whereas the two dashed lines represent the interquartile range. \label{fig:distr_validation}}
\end{figure}

\section[\appendixname~\thesection]{Validation details}
\label{sec:appendix_validation}

As described in Section \ref{sec:methods_validation}, the new method for network construction implemented in NEST GPU needs an in-depth analysis for validating the new version against the previous version of the library. To verify the quality of the results we collect the spiking activity of the neuron populations of the cortical microcircuit model, and compute three distributions of the spiking activity to be compared, i.e., the average firing rate of the populations, the coefficient of variation of inter-spike-intervals (CV ISI) and the pairwise Pearson correlation of the spike trains for each population.
The simulations are performed using a time step of $0.1$\,ms and $500$\,ms of network dynamics are simulated before recording the spiking activity to avoid transients. Then, the spiking activity of the subsequent $600$\,s of network dynamics is recorded to compute the distributions. As shown in \cite{Dasbach2021}, this large amount of biological time to be simulated is needed to let the activity statistics converge, and thus to be able to distinguish the statistic of the activity from random processes. Regarding the performance of such simulations, the real-time factor of NEST GPU with enabled spike recording has only a $1.5\%$ increase with respect to the performance shown in Figure \ref{fig:microcircuit_nestgpu_comparison}.
Figure \ref{fig:distr_validation} shows the violin plots of the distributions obtained with the \texttt{seaborn.violinplot} function of the Seaborn \citep{seaborn} Python library. The function computes smoothed distribution through the Kernel Density Estimation method \citep{Rosenblatt1956, Parzen1962} with Gaussian kernel, with bandwidth optimized using the Silverman method \citep{silverman86}.

\begin{figure}[b]
\centering
\includegraphics[width=\textwidth]{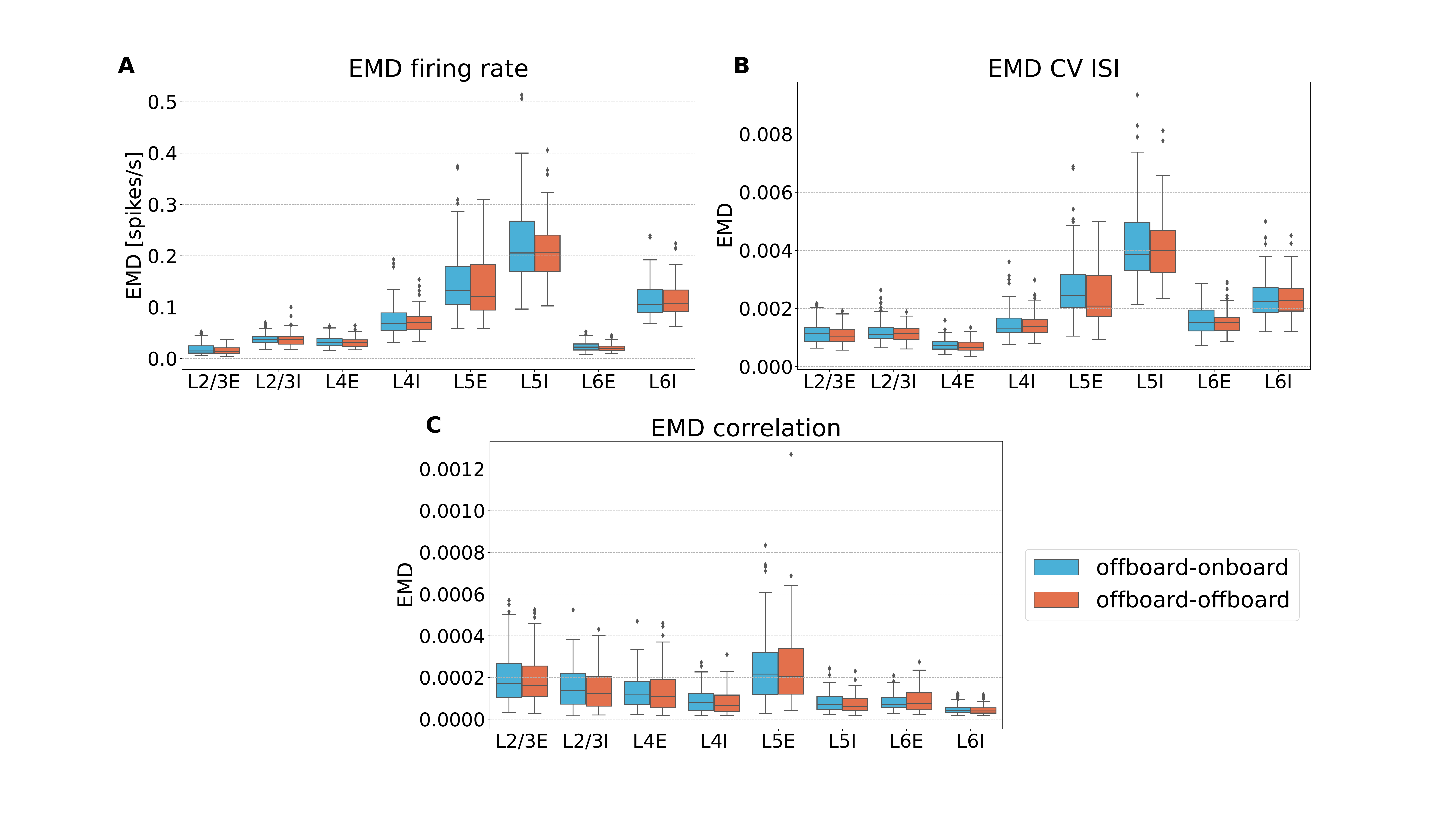}
\includegraphics[width=\textwidth]{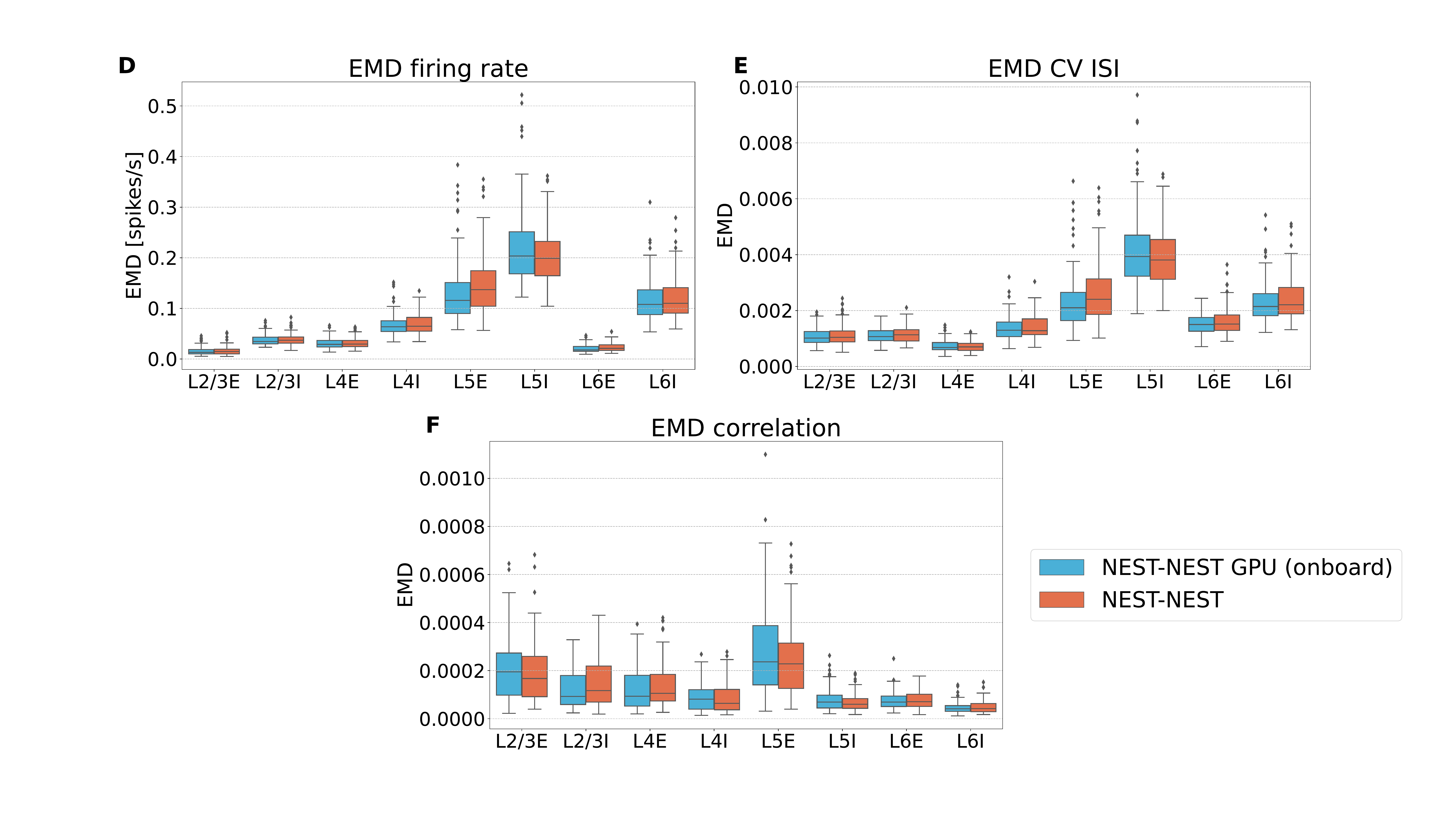}
\caption{
Box plots of the Earth Mover's Distance comparing side by side firing rate \textbf{(A)}, CV ISI \textbf{(B)} and Pearson correlation \textbf{(C)} of the two versions of NEST GPU (sky blue boxes, left) and the previous version of NEST GPU using different seeds (orange boxes, right). Panels \textbf{(D, E, F)} are the same as \textbf{(A, B, C)} but distributions of NEST GPU (onboard) and NEST 3.3 are compared. In particular, the comparison between the different simulator is represented by the sky blue boxes on the left, whereas the comparison between two sets of NEST simulations is depicted with the orange boxes. Central dashed line represents the median of the distributions, whereas the two dashed lines represent the interquartile range.
\label{fig:emd_validation}}
\end{figure}

The distributions obtained with the two versions of NEST GPU are visually indistinguishable; distributions of the CPU simulator (version 3.3) are likewise indistinguishable, as previously demonstrated in \cite{Golosio2021} for the comparison between the previous version of NEST and NeuronGPU, the prototype library of NEST GPU. Additionally, to quantitatively evaluate the difference between the different versions of NEST GPU we compute the Earth Mover's Distance (EMD) between pairs of distributions using the \texttt{scipy.stats.wasserstein\_distance} of the SciPy library \citep{SciPy}. More details on this method can be found in \cite{Tiddia2022}. We simulate sets of 100 simulations changing the seed for random number generation. The sets of simulations for the two versions of the NEST GPU library are thus pairwise compared, obtaining for each distribution and each population of the model a set of 100 values of EMD, evaluating the difference between the distributions of the two versions of NEST GPU (\textit{offboard}-\textit{onboard}). Furthermore, we compute an additional set of simulations for the previous version of NEST GPU, to be compared with the other set of the same version (\textit{offboard}-\textit{offboard}). This way, we can evaluate the differences that can arise using the same simulator with different seeds for random number generation and compare it with the differences obtained by comparing the two different versions of NEST GPU. Additionally, we performed the same validation to compare NEST and NEST GPU to have a quantitative comparison between the most recent versions of the two simulators, i.e., \textit{NEST-NEST GPU (onboard)} and \textit{NEST-NEST}. Figure \ref{fig:emd_validation} shows the EMD box plots for all the distributions computed and for all the populations.\\
Comparing the box plots in panels A-C reveals very similar distributions in EMD for the two comparison, meaning that the variability we measure from the comparing the two versions is compatible to the one that we have by employing the previous version of NEST GPU using different seeds, ergo the new method does not add variability with respect to simulating the model with the previous version of NEST GPU using different seeds. Similar conclusions can be derived from the comparison between NEST and NEST GPU \textit{(onboard)} (see panels D-F).\\
% Don't remove:
% This means that NEST GPU main tooks ~920 minutes to simulate 100 simulations of 600s of biological time, whereas NEST GPU conn takes 867 minutes to do that, which means that there is a difference of ~50 minutes due to the optimization of the simulation pase (i.e. the nested loop) done in NEST GPU onboard
As mentioned before, the real-time factor of NEST GPU marginally increased because of the activation of the spike recording. The overall simulation time of a set of 100 simulations using the novel method for network construction took around 868 minutes, with less than one minute dedicated to network construction (more precisely, the average time is $0.53$\,s for a single simulation). A set of simulations obtained using the old method for network construction took around 1020 minutes, with around $100$ minutes of them related to the network construction phase. This represents a reduction of the network construction time of around 116 times with respect to the previous network construction method.

\begin{figure}[H]
\centering
\includegraphics[width=0.9\textwidth]{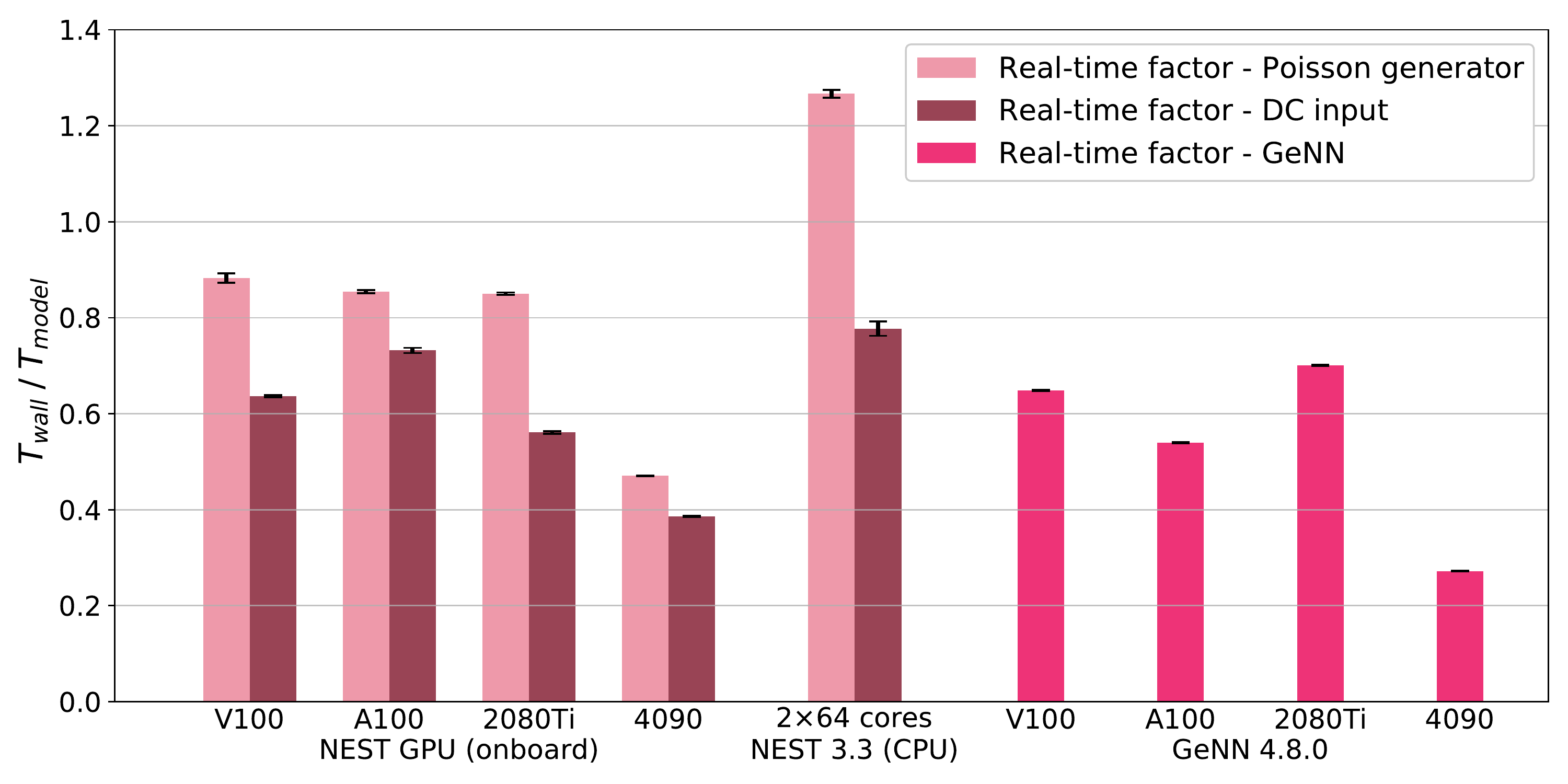}
\caption{Real-time factor, defined as $T_{\text{wall}}/T_{\text{model}}$, of cortical microcircuit model simulations for NEST GPU \textit{(onboard)}, NEST and GeNN. The biological model time we use to compute the real-time factor is $T_{\text{model}}=10$\,s, simulated driving the external stimulation using Poisson spike generators (left bars, pink) or DC input (right bars, dark red). GeNN (magenta bars) employs a different approach for simulating external stimuli.
Error bars show the standard deviation of the simulation phase over ten simulations using different random seeds.
\label{fig:microcircuit_rtf}}
\end{figure}

\section[\appendixname~\thesection]{Additional data for cortical microcircuit simulations}
\label{sec:appendix_microcircuit}

Analog to Figure \ref{fig:microcircuit_nestgpu_comparison}B, we show in Figure \ref{fig:microcircuit_rtf} the real-time factor for simulations run with the CPU version of NEST and GeNN.

For the CPU version of NEST, \cite{Kurth2022} demonstrate a smaller real-time factor for simulations of the cortical microcircuit model with DC input compared to our results in Figure \ref{fig:microcircuit_rtf}, which is likely due to a different version of the simulation code. We also employ a different parallelization strategy to optimize the real-time factor with the recent release NEST 3.3 on a compute node of the JURECA-DC cluster (i.e., 8 MPI processes each running 16 threads, as in \cite{Albers2022} who obtained similar results with NEST 3.0).

\section[\appendixname~\thesection]{Additional data for the two-population network simulations}
\label{sec:appendix_izhikevich}
\begin{figure}[H]
\centering
\includegraphics[width=0.75\textwidth]{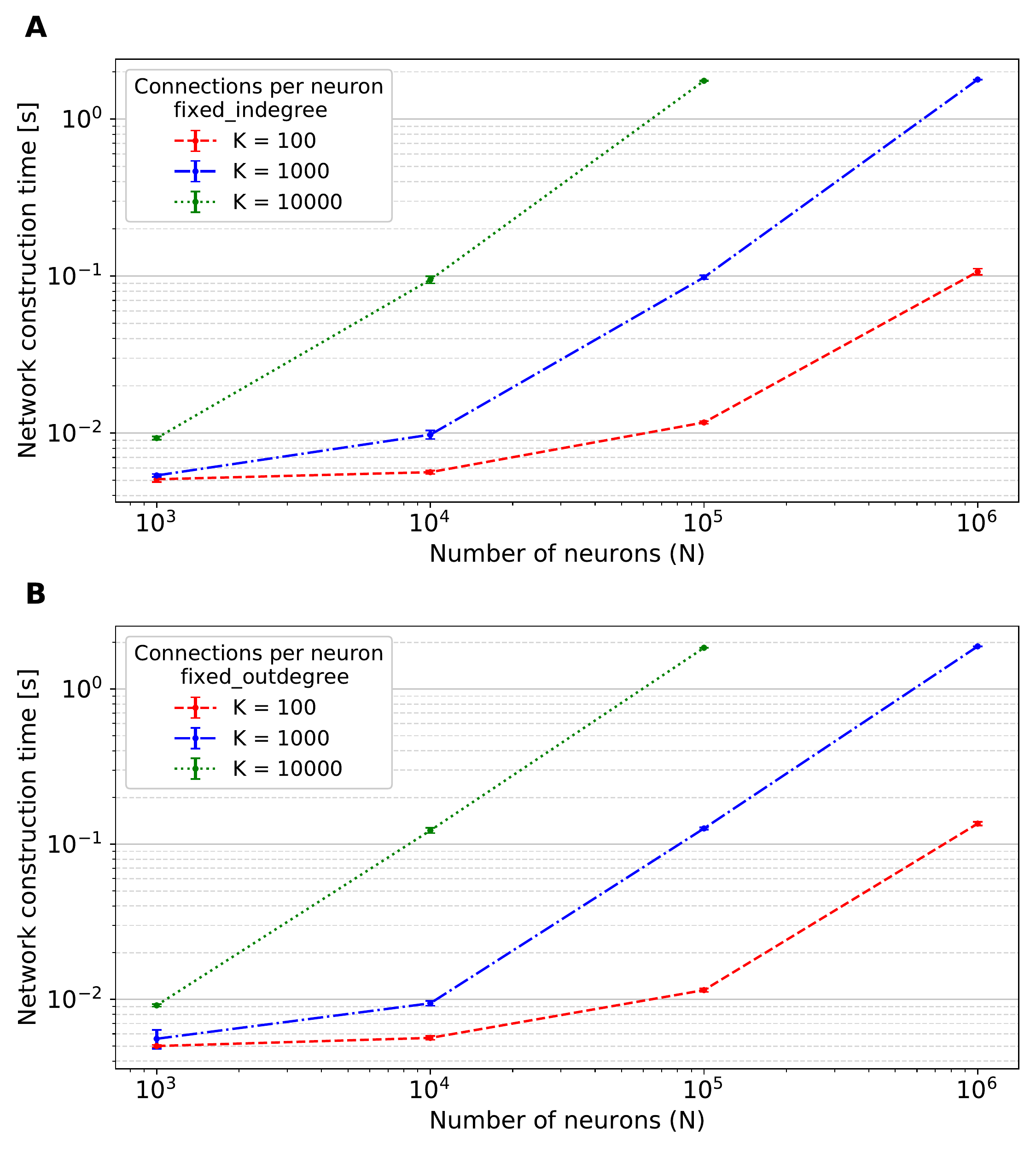}
\caption{Network construction time of the two-population network with $N$ total neurons and $K$ connections per neuron using different connection rules. \textbf{(A)}~Performance obtained using the \texttt{fixed\_indegree} connection rule, i.e., each neuron of the network has an in-degree of $K$. \textbf{(B)}~Performance obtained using the \texttt{fixed\_outdegree} connection rule, i.e., each neuron of the network has $K$ out-degrees. The value of network construction time for the network with $10^6$ neurons and $10^4$ connections per neuron is not shown because of lack of GPU memory. Error bars indicate the standard deviation of the performance across 10 simulations using different seeds.
\label{fig:fixed_indegree_outdegree}}
\end{figure}

Figure \ref{fig:fixed_tot_num} shows the network construction time of the two-population network using the \texttt{fixed\_total\_number} connection rule. In Figure \ref{fig:fixed_indegree_outdegree}, we provide the corresponding data for the \texttt{fixed\_indegree} and the \texttt{fixed\_outdegree} rules.

\end{document}